\documentclass[twocolumn]{aastex}
\usepackage{amsmath,esint}
\usepackage{url,aasmacros}
\usepackage{natbib}
\usepackage{soul, CJK}
\usepackage{multirow}
\usepackage{tablefootnote}

\usepackage{bm}
\usepackage{xspace}
\usepackage{color}
\usepackage{enumitem}
\usepackage{booktabs}

\usepackage[utf8]{inputenc}

\shorttitle{Kilonova Rate Constrained by ZTF}
\shortauthors{Andreoni et al.}

\begin{document}

\title{	
Constraining the Kilonova Rate with Zwicky Transient Facility Searches \\Independent of Gravitational Wave and Short Gamma-ray Burst Triggers 
}

\author[0000-0002-8977-1498]{Igor~Andreoni}
\email{andreoni@caltech.edu}
\affiliation{Division of Physics, Mathematics and Astronomy, California Institute of Technology, Pasadena, CA 91125, USA}

\author[0000-0002-7252-3877]{Erik C. Kool}
\affiliation{The Oskar Klein Centre, Department of Astronomy, Stockholm University, AlbaNova, SE-106 91 Stockholm, Sweden}

\author{Ana Sagu{\'e}s Carracedo}
\affiliation{The Oskar Klein Centre, Department of Physics, Stockholm University, AlbaNova, SE-106 91 Stockholm, Sweden}

\author{Mansi M. Kasliwal}
\affiliation{Division of Physics, Mathematics and Astronomy, California Institute of Technology, Pasadena, CA 91125, USA}

\author[0000-0002-8255-5127]{Mattia Bulla}
\affiliation{Nordita, KTH Royal Institute of Technology and Stockholm University, Roslagstullsbacken 23, SE-106 91 Stockholm, Sweden}

\author[0000-0002-2184-6430]{Tom{\'a}s Ahumada}
\affiliation{Department of Astronomy, University of Maryland, College Park, MD 20742, USA}

\author[0000-0002-8262-2924]{Michael W. Coughlin}
\affil{School of Physics and Astronomy, University of Minnesota, Minneapolis, Minnesota 55455, USA}

\author[0000-0003-3768-7515]{Shreya Anand}
\affiliation{Division of Physics, Mathematics and Astronomy, California Institute of Technology, Pasadena, CA 91125, USA}

\author{Jesper Sollerman}
\affiliation{The Oskar Klein Centre, Department of Astronomy, Stockholm University, AlbaNova, SE-106 91 Stockholm, Sweden}

\author{Ariel Goobar}
\affil{The Oskar Klein Centre, Department of Physics, Stockholm University, AlbaNova, SE-106 91 Stockholm, Sweden}

\author{David L. Kaplan}
\affiliation{Center for Gravitation, Cosmology, and Astrophysics, Department of Physics, University of Wisconsin-Milwaukee, P.O. Box 413,
Milwaukee, WI 53201, USA}

\author{Tegan T. Loveridge}
\affil{Division of Physics, Mathematics and Astronomy, California Institute of Technology, Pasadena, CA 91125, USA}

\author{Viraj Karambelkar}
\affil{Division of Physics, Mathematics and Astronomy, California Institute of Technology, Pasadena, CA 91125, USA}

\author[0000-0001-5703-2108]{Jeff Cooke}
\affil{Australian Research Council Centre of Excellence for Gravitational Wave Discovery (OzGrav), Swinburne University of Technology,
Hawthorn, VIC, 3122, Australia}
\affil{Centre for Astrophysics and Supercomputing, Swinburne University of Technology, Hawthorn, VIC, 3122, Australia}

\author{Ashot Bagdasaryan}
\affil{Division of Physics, Mathematics and Astronomy, California Institute of Technology, Pasadena, CA 91125, USA}

\author[0000-0001-8018-5348]{Eric C. Bellm}
\affiliation{DIRAC Institute, Department of Astronomy, University of Washington, 3910 15th Avenue NE, Seattle, WA 98195, USA} 

\author{S. Bradley Cenko}
\affil{Astrophysics Science Division, NASA Goddard Space Flight Center, Mail Code 661, Greenbelt, MD 20771, USA}
\affil{Joint Space-Science Institute, University of Maryland, College Park, MD 20742, USA}

\author[0000-0002-6877-7655]{David O. Cook}
\affiliation{IPAC, California Institute of Technology, 1200 E. California Blvd, Pasadena, CA 91125, USA}

\author{Kishalay De}
\affil{Division of Physics, Mathematics and Astronomy, California Institute of Technology, Pasadena, CA 91125, USA}

\author[0000-0002-5884-7867]{Richard Dekany}
\affiliation{Caltech Optical Observatories, California Institute of Technology, Pasadena, CA 91125, USA}

\author{Alexandre Delacroix}
\affiliation{Caltech Optical Observatories, California Institute of Technology, Pasadena, CA 91125, USA}

\author{Andrew Drake}
\affil{Division of Physics, Mathematics and Astronomy, California Institute of Technology, Pasadena, CA 91125, USA}

\author[0000-0001-5060-8733]{Dmitry A. Duev}
\affil{Division of Physics, Mathematics and Astronomy, California Institute of Technology, Pasadena, CA 91125, USA}

\author[0000-0002-4223-103X]{Christoffer Fremling}
\affil{Division of Physics, Mathematics and Astronomy, California Institute of Technology, Pasadena, CA 91125, USA}

\author[0000-0001-8205-2506]{V. Zach Golkhou}
\affiliation{DIRAC Institute, Department of Astronomy, University of Washington, 3910 15th Avenue NE, Seattle, WA 98195, USA} 
\affiliation{The eScience Institute, University of Washington, Seattle, WA 98195, USA}

\author{Matthew J. Graham}
\affiliation{Division of Physics, Mathematics and Astronomy, California Institute of Technology, Pasadena, CA 91125, USA}

\author{David Hale}
\affiliation{Caltech Optical Observatories, California Institute of Technology, Pasadena, CA 91125, USA}

\author{S. R. Kulkarni}
\affiliation{Division of Physics, Mathematics and Astronomy, California Institute of Technology, Pasadena, CA 91125, USA}

\author[0000-0002-6540-1484]{Thomas Kupfer}
\affiliation{Kavli Institute for Theoretical Physics, University of California, Santa Barbara, CA 93106, USA}

\author[0000-0003-2451-5482]{Russ R. Laher}
\affiliation{IPAC, California Institute of Technology, 1200 E. California Blvd, Pasadena, CA 91125, USA}

\author[0000-0003-2242-0244]{Ashish A. Mahabal}
\affiliation{Division of Physics, Mathematics and Astronomy, California Institute of Technology, Pasadena, CA 91125, USA}

\author[0000-0002-8532-9395]{Frank J. Masci}
\affiliation{IPAC, California Institute of Technology, 1200 E. California Blvd, Pasadena, CA 91125, USA}

\author[0000-0001-7648-4142]{Ben Rusholme}
\affiliation{IPAC, California Institute of Technology, 1200 E. California Blvd, Pasadena, CA 91125, USA}

\author[0000-0001-7062-9726]{Roger M. Smith}
\affiliation{Caltech Optical Observatories, California Institute of Technology, Pasadena, CA 91125, USA}

\author{Anastasios Tzanidakis}
\affil{Division of Physics, Mathematics and Astronomy, California Institute of Technology, Pasadena, CA 91125, USA}

\author[0000-0003-4131-173X]{Angela Van Sistine}
\affiliation{Center for Gravitation, Cosmology, and Astrophysics, Department of Physics, University of Wisconsin-Milwaukee, P.O. Box 413,
Milwaukee, WI 53201, USA}

\author[0000-0001-6747-8509]{Yuhan Yao}
\affil{Division of Physics, Mathematics and Astronomy, California Institute of Technology, Pasadena, CA 91125, USA}

\begin{abstract}
The first binary neutron star merger, GW170817,  was accompanied by a radioactivity-powered optical/infrared transient called a kilonova. To date, no compelling kilonova has been found during optical surveys of the sky, independent of gravitational-wave triggers. In this work, we searched the first 23 months of the Zwicky Transient Facility (ZTF) data stream for candidate kilonovae in the form of rapidly evolving transients. We combined ZTF alert queries with forced point-spread-function photometry and nightly flux stacking to increase our sensitivity to faint and fast transients. Automatic queries yielded $>11,200$ candidates, 24 of which passed quality checks and strict selection criteria based on a grid of kilonova models tailored for both binary neutron star and neutron star--black hole mergers. None of the candidates in our sample was deemed a possible kilonova after thorough vetting, catalog cross-matching, and study of their color evolution. The sources that passed our selection criteria are dominated by Galactic cataclysmic variables. In addition, we identified two fast transients at high Galactic latitude, one of which is the confirmed afterglow of long-duration GRB~190106A, and the other is a possible cosmological afterglow. Using a survey simulation code, we constrained the kilonova rate for a range of models including top-hat and linearly decaying light curves and synthetic light curves obtained with radiative transfer simulations. For prototypical GW170817-like kilonovae, we constrain the rate to be $R < 1775$\,Gpc$^{-3}$\,yr$^{-1}$ at 95\% confidence level by requiring at least 2 high-significance detections. By assuming a population of kilonovae with the same geometry and composition of GW170817 observed under a uniform viewing angle distribution, we obtained a constraint on the rate of $R < 4029$\,Gpc$^{-3}$\,yr$^{-1}$.

\end{abstract}

\section{Introduction}
\label{sec: intro}

The study of the dynamic sky took a twist when multi-messenger discoveries became reality with the discovery of astronomical transients associated with neutrinos \citep{Hirata1987, IceCube2018Sci, Stein2020arXiv} or gravitational waves \citep[GWs;][]{Abbott2017MMA}.
The identification of an electromagnetic (EM) counterpart to the GW event GW170817 \citep{Abbott2017MMA} spectacularly confirmed the predictions that short gamma-ray bursts \citep[GRBs; e.g.][]{Blinnikov1984,Paczynski1986,Eichler1989a, Narayan1992,Fong&Berger2013} and fast optical/near-infrared transients called ``kilonovae" or ``macronovae" \citep[KNe; e.g.][]{Li1998,Kulkarni2005,Rosswog2005, Metzger2010a, Metzger2012, Barnes2013, Tanaka2013} can be generated by binary neutron star (BNS) or neutron star--black hole (NSBH) mergers \citep[see also][]{Foucart:2012nc, Hotokezaka2013}. 
Launched by neutron star disruption, neutron-rich ejecta host rapid neutron capture (r-process) nucleosynthesis, leading to the production of heavy elements. The radioactive decay of such unstable heavy nuclei powers the optical/infrared KN. Recent reviews on EM emission from neutron star mergers can be found, for example, in \cite{Metzger2019LRR, Nakar2019arXiv}.

The observation of optical/infrared KNe is particularly valuable when it happens concurrently with the discovery of a compact binary merger in GWs. The nearby BNS merger GW170817 was confidently associated with the optical transient AT2017gfo \citep{Coulter2017} and the multi-messenger discovery has led to hundreds of studies addressing, for example, astrophysics of energetic phenomena \citep[e.g.,][]{Kasliwal2017, DAvanzo2018X, Margutti2018, Beniamini2019MNRAS, SharanSalafia2020arXiv}, fundamental physics \citep[e.g.,][]{Abbott2019PhRvX, Coughlin2019MNRAS}, and cosmology \citep[e.g.,][]{Abbott2017cosmology, Hotokezaka2019NatAs}. It is relevant for this work to note that even the optical/infrared photometric data alone carried evidence of heavy-element nucleosynthesis in the GW170817 merger ejecta \citep{Andreoni2017gw, Arcavi2017GW, Chornock2017, Cowperthwaite2017, Diaz2017, Drout2017Sci, Evans2017Sci, Kasliwal2017, Pian2017Nat, Smartt2017, Tanvir2017, Troja2017Nat, Shappee2017, Utsumi2017, Valenti2017}.

The discovery of more KNe in optical surveys would allow us to estimate the production of heavy elements during neutron star mergers and better constrain the neutron star merger rate (see \S\ref{sec: rates}), therefore providing tools to tackle open questions such as whether neutron star mergers are ``the site" or just ``a site" for heavy element production in the Universe. This topic is discussed, for example, by \cite{Rosswog2018, Kasliwal2019spitzer, Siegel2019Nat}.
The possibility of using KNe as standardizable candles for cosmology (without any GW information) is also an intriguing perspective \citep{Coughlin2020KNstandard} that can be tested when a significant population of KNe is unveiled. 

Observationally, KNe can show an early blue component likely brighter towards polar angles for the first 2--3 days after merger, followed by a redder component generated from the tidal dynamical ejecta and post-merger ejecta \citep[e.g.,][]{Cowperthwaite2017, Kasen2017Nat, Kawaguchi2018, Bulla2019NatAs}. Regardless of the viewing angle, KNe are expected to appear dimmer and fade more quickly relative most supernovae at optical wavelengths. Supernovae typically peak at absolute magnitudes between $\sim -17$ and $\sim -19$ and, in most cases, they fade at a rate slower than 0.3\,mag\,day$^{-1}$. AT2017gfo faded at a rate $>0.5$\,mag\,day$^{-1}$ in optical bands and it was measured to have an absolute magnitude of $M_r \sim -16$ shortly after peak. Such characteristics make it more difficult to discover KNe relative to supernovae, that are brighter at peak and remain luminous for months. Moreover, recent GW observations indicate a rate for BNS mergers of 250--2810\,Gpc$^{-3}$\,yr$^{-1}$ \citep{Abbott2020GW190425discovery}, which suggests that at most few events per year are expected to occur at distances close enough for KNe to be detectable in 1-m class optical surveys.  In comparison, the rate of unobscured supernovae in the nearby Universe is $\sim 28.7 \times 10^3$\,Gpc$^{-3}$\,yr$^{-1}$ \citep{Smartt2009b}.

Several KN candidates were found during the follow-up of short GRBs \citep{Perley2009, Tanvir2013a, Berger2013k, Gao2015, Jin2015, Yang2015, Jin2016, Troja2018KN, Jin2020NatAs} as optical/IR excesses on top of the GRB afterglow and one was found during the follow-up of GW170817/GRB170817A \citep[e.g.,][]{Coulter2017}. Although a marginal candidate was recently identified \citep{McBrien2020arXiv}, no confirmed KN was found during optical surveys independently of GW triggers, to date. 

Searches in optical time-domain survey datasets are complementary to GW follow-up searches. The third Advanced LIGO/Virgo observing run (O3) took place entirely within the time frame in which data used in this work were taken and it was suspended without yielding any optical (or multi-wavelength) counterpart \citep{Ackley2020S190814bvarXiv, Andreoni2019S190510g, Coughlin2019GW190425, Goldstein2019S190426c, Gomez2019S190814bv,
Hosseinzadeh2019S190925z, Lundquist2019, Andreoni2020S190814bv, Antier2020arxiv, Gompertz2020GOTOarXiv, Vieira2020arxiv, Watson2020}. In addition, it is possible that neutron star mergers occurred in the nearby Universe during O3 and were missed by the GW detectors because of instrument downtime and their unisotropic antenna pattern. 

Motivated by this, we use the Zwicky Transient Facility \citep[ZTF;][]{Bellm2019ZTF, Graham2019ZTF} at Palomar Observatory, which surveys the northern sky primarily in $g$- and $r$-band ($i$-band being occasionally used for specific projects) with an instantaneous field of view of 47\,deg$^2$, a pixel scale of $1''$\,pixel$^{-1}$, and a typical limiting magnitude of $r\sim20.5$ in 30-s exposure time, to search for serendipitous KNe. This work is based on 239,754 images acquired during the first 23 months of ZTF operations.

The paper is organized as follows: methods used to search the ZTF alert database, to perform photometry, and to select KN candidates are described in \S\ref{sec: methods}; the results of our searches are presented in \S\ref{sec: results}; implications of our results for KN rates are discussed in \S\ref{sec: rates}. \S\ref{sec: conclusion} completes the manuscript with a short summary and concluding remarks. A \cite{Planck2015cosmo} cosmology is used throughout the paper.

\section{Methods}
\label{sec: methods}

In this section, we describe the dataset that we searched for KNe, the data mining techniques that we adopted, and our novel scheme to vet candidates.

\subsection{Dataset}
\label{sec: dataset}

ZTF conducts a public Survey of the Northern Sky for 40\% of the time allocation. Public data are acquired as $g$, $r$ filter pairs with a preferred cadence of 3 nights, with some fields in the Galactic plane or in {\it Transiting Exoplanet Survey Satellite} \citep[{\it TESS};][]{Ricker2015TESS} fields imaged with nightly cadence \citep[for more information on ZTF goals and planning see][]{Bellm2019scheduler}. The rest of the time allocation is divided between ZTF Partnership and Caltech surveys, usually observing a smaller sky area (including the Galactic plane) at higher cadence. Data from all programs were used in our analysis. Images are processed with the ZTF real-time reduction and image subtraction pipeline at the Infrared Processing \& Analysis Center \citep{Masci2019ZTF}, using the \texttt{ZOGY} algorithm for the image subtraction \citep{Zackay2016}. Upon source detection, alert packets are generated and serialized in \texttt{Apache Avro} format, and finally distributed to community brokers \citep{Patterson2019ZTF}.

\begin{figure}[t]
\centering
\includegraphics[width=\columnwidth]{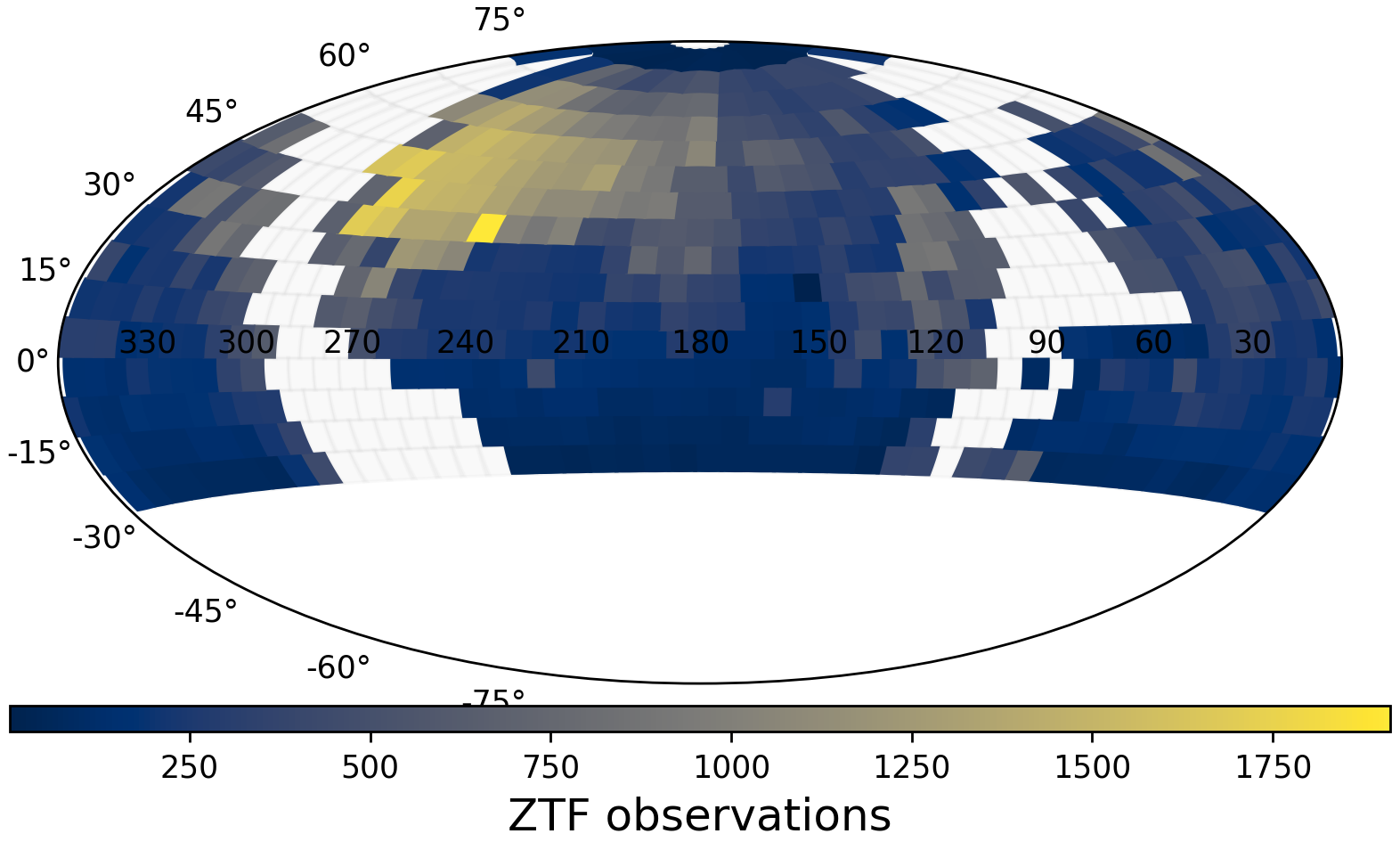}
\caption{All-sky view of the ZTF fields searched in this work created with \texttt{ztfquery} \citep{mickael_rigault_2018_1345222}, in Equatorial coordinates and  ``hammer" projection. ZTF fields were selected to minimize the Galactic foreground by setting a threshold of $E(B-V) <$ 0.3\,mag at their central coordinates \citep{Planck2014dust}. }
\label{fig: fields}
\end{figure}

This work aims to discover extragalactic transients, so we limit the search for KNe to 514 fields (47\,deg$^2$ each) with low Galactic extinction by requiring $E(B-V) < 0.3$\,mag at the central coordinates for the field based on \cite{Planck2014dust} dust maps. The central coordinates of ZTF fields belong to two fixed grids, where the secondary grid is shifted from the primary grid in order to cover the chip gaps. We only take fields in the well-sampled ZTF primary grid, given that the secondary grid has significantly fewer observations. The effective area that we probed extended across $\sim$\,24,150\,deg$^2$ (Figure~\ref{fig: fields}). 
The resulting dataset utilizes 239,754 images, 89,961 of which were acquired in $g$-band, 137,403 in $r$-band, and 12,390 in $i$-band.

\subsection{Alert database query}
\label{sec: query}

The ZTF alert database was mined by querying the \texttt{Kowalski}\footnote{\url{https://github.com/dmitryduev/kowalski}} broker at Caltech \citep{Duev2019} in \texttt{mongodb} and \texttt{Python}. ZTF alerts are issued when a source is detected with $\geq 5\sigma$ significance, however the number of detections reported within the \texttt{Avro} packets (\texttt{ndethist} key) takes into account past detections with a lower significance, ranging between $3-5$\,$\sigma$.  When querying the database, we excluded those candidates that were:

\begin{itemize}
    \item classified with a small real/bogus score of \texttt{rb}$ < 0.5$ and deep learning scores of \texttt{braai}$< 0.8$ and \texttt{drb}$< 0.8$, where available \citep{Duev2019}; 
    \item located $<10''$ from known Solar System objects;
    \item located $< 15''$ from bright ($r < 15$) stars that often produce artificial spikes and ghost sources in their surroundings;
    \item detected as ``negative" flux in the image subtraction, the source having become fainter in the science image than in the reference image;
    \item coincident within $2''$ radius with likely stellar sources (\texttt{sgscore}$> 0.5$) found in Pan-STARRS images based on the star/galaxy separation algorithm developed by \cite{Miller2017}.
\end{itemize}

We organized the experiment in two parts in order to conduct a thorough analysis on a large but manageable dataset:

\begin{enumerate}

    \item an {\it all-sky} search in the selected fields, requiring at least 3 detections in the alerts and at most 12 days time lag between the first and last detection. Such a time lag would be sufficient to observe AT2017gfo appear and fade away in our data (limiting magnitude $\sim 20.5$) at 40\,Mpc, although KNe are expected to be found at larger distances, remaining bright for even less time above our detection threshold. In addition, bright ($r < 18.5$\,mag) nearby KNe should have been identified in near real-time during the systematic spectroscopic classification effort of the ZTF Bright Transient Survey \citep{Fremling2020RCF}, or part of ZTF collaborative programs aiming at young/fast transient discovery \citep[e.g.,][]{Ho2020koala, Yao2019}. The query returned 10,419 candidates.
    
    \item a {\it galaxy-targeted} search, requiring at least 2 detections in the alerts and at most 6 days lag between the first and last detection, with a positive crossmatch within a 100\,kpc radius (see \S\ref{subsec: all-sky}) with galaxies in the Census of the Local Universe \citep[CLU;][]{Cook2017CLU} catalog further than 10\,Mpc. ZTF should be sensitive to faint $M_r\sim -9.5$ novae at 10\,Mpc, thus KNe at that distance (or closer) are expected to appear as bright transients that our all-sky search should be able to detect easily. Faint ($r < 20$\,mag) KNe detected several times in known galaxies closer than 200\,Mpc could have been found in near real-time and classified during the Caltech volume-limited ``CLU experiment" \citep{De2020CLUarXiv}, but this experiment has not yielded any KN candidate, yet. This query returned 1,569 candidates, 695 of which were possibly associated with galaxies $>40$\,Mpc away.

\end{enumerate}

Together, these searches yielded 11,202 unique candidates (10,419 from the all-sky search and 1,569 from the galaxy-targeted search, with an overlap of 786 candidates). We removed from the sample 79 sources with a counterpart in the {\it Gaia} DR2 catalog (separation $<1.5''$) with a parallax measured to be inconsistent with $0 \pm 1.081$\,mas, leaving 11,123 candidates to be further analyzed. This ``stellarity" threshold was determined from the study of AllWISE catalog quasars present in the {\it Gaia} dataset \citep{Luri2018}.

\subsection{Photometry and candidate selection criteria}
\label{sec: photometry and detection criteria}

Forced point spread function (PSF) photometry was performed at the location of each candidate using \texttt{ForcePhotZTF} \citep{Yao2019}. The median coordinates recorded in the available alerts was used to improve the location accuracy. A signal-to-noise ratio of S/N$=3$ served as threshold for the photometric detection of a source. 

\begin{figure}[t]
\centering
\includegraphics[width=\columnwidth]{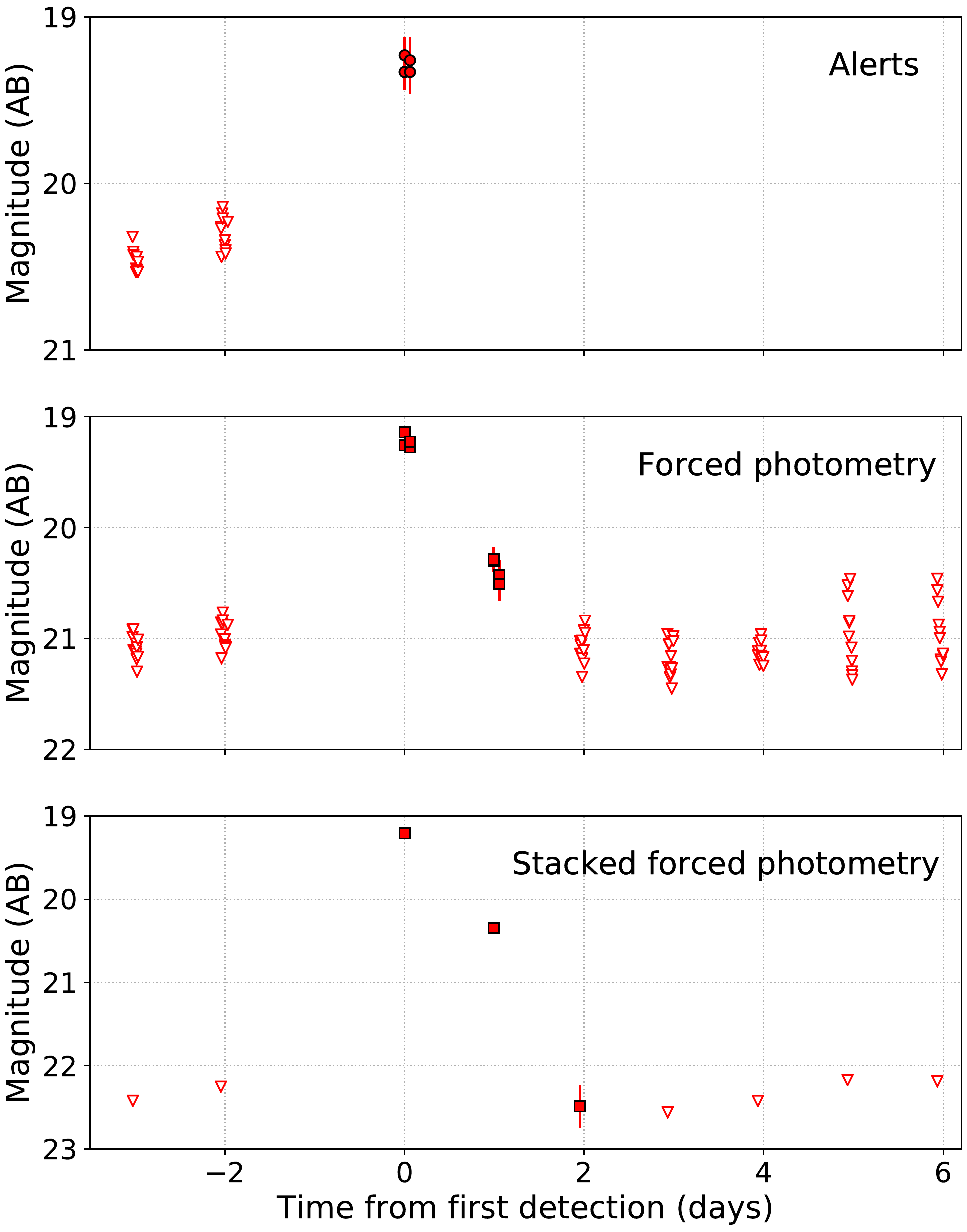}
\caption{Light curve of the candidate ZTF19abcpiag in $r$-band built with ZTF alerts ({\it top}), forced photometry {\it center}), and nightly-stacked forced photometry ({\it bottom}). The light curve based only on ZTF alerts is shallower and offers this transient detection on 1 night only.  Single-epoch photometry with alerts or forced photometry could reveal intra-night variability, while stacked forced photometry allowed us to monitor the evolution of faint transients for longer time. In this project we used a combination of these photometric techniques to search for KNe in ZTF data.}
\label{fig: example photometry}
\end{figure}

First, we ran forced photometry on images taken starting a week before the first detection and until 2 weeks past the last detection of each candidate, in order to be less biased by spurious detections consistent with Gaussian noise in several thousands of images by considering the whole survey.
Then, we stacked the forced photometry data points nightly in flux space to increase our sensitivity to faint sources. 
Finally, we ran forced photometry at the location of selected candidates for the whole duration of the survey and we rejected those candidates showing repeated activity that was not significant enough to be recorded in ZTF alerts.
An example of how forced photometry and nightly stacked photometry can help reveal the behavior of faint transients is shown in Figure\,\ref{fig: example photometry}. We note that public ZTF images acquired after Public Data Release 2 on 2019 December 11 were not yet released for forced photometry when this work was carried out.

\begin{figure*}[t]
\centering
\includegraphics[width=2\columnwidth]{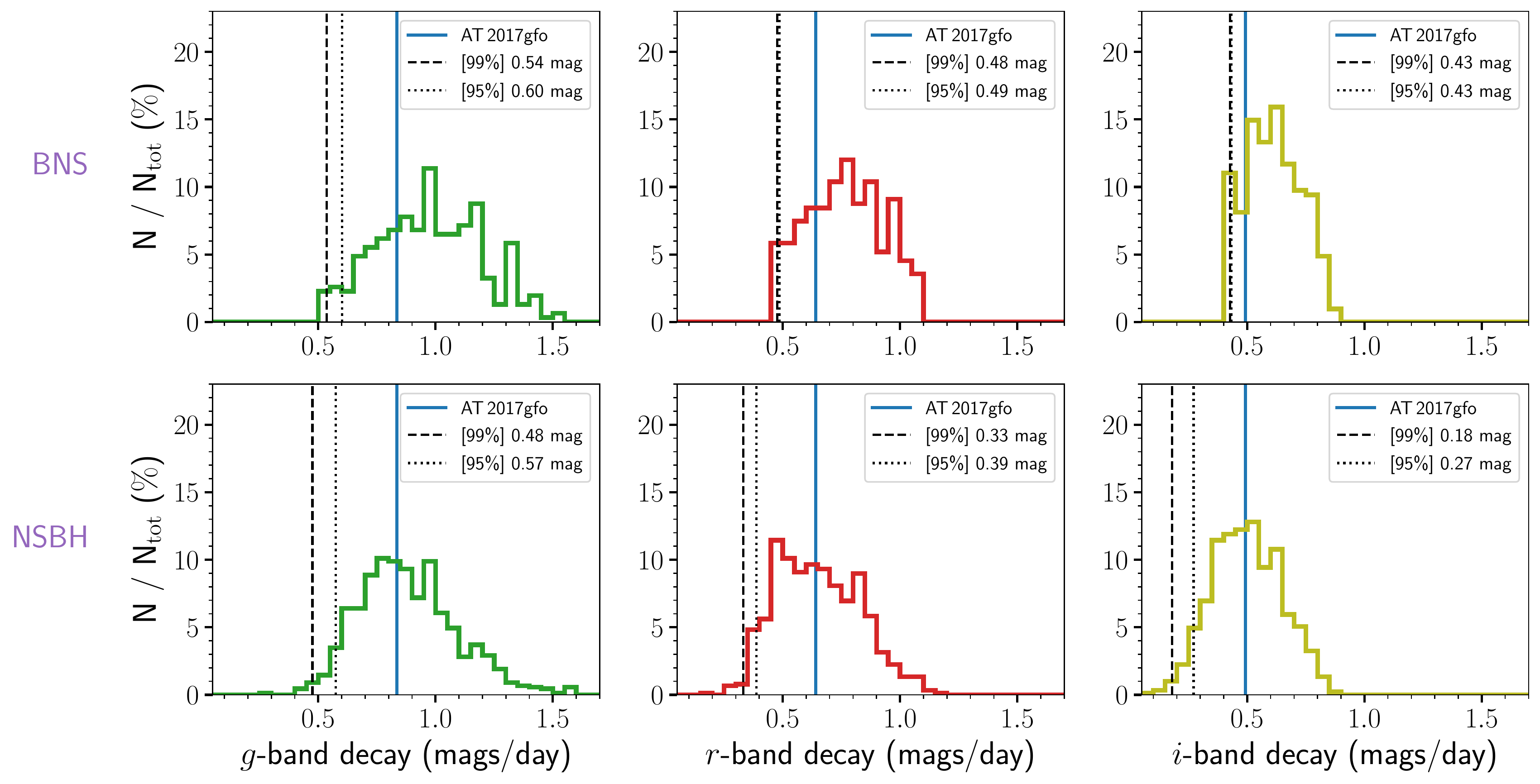}
\caption{A grid of models tailored for BNS ({\it upper} panel) and NSBH ({\it lower} panel) mergers was used to determine suitable thresholds for KN candidate selection. Decay rates from peak to +6\,d later are shown in $g$-, $r$- and $i$-band from left to right. The lower 99\% and 95\% thresholds for each distribution are marked by dashed and dot-dashed vertical lines, respectively, and reported in the legend of each panel. Blue vertical lines mark the decay rates for AT2017gfo, the KN of GW170817.}
\label{fig: threshold}
\end{figure*}

KNe are expected to fade rapidly, as AT2017gfo clearly confirmed. 
We automatically reject all those candidates that do not show any $>2\sigma$ evolution for 6, 8, and 10 days from the first detection in $g$-, $r$-, and $i$-band respectively.
Forced photometry and nightly stacking increased the measured duration of some candidates because they allowed us to recover additional faint data points. Thus we apply another filter on stacked light curves to reject those candidates with total time span $>14$ days and with time lag between first and last detection larger than 10, 12, 14 days in $g$-, $r$-, and $i$-band, respectively.

Where possible, we use two different linear fits before and after the brightest point to measure the rise and fade rates. In each band, we required a minimum time baseline of 3 hours between first and last detection for the data points to be fit. If the brightest point of the light curve occurred within 3 hours of the first or the last detection, a single linear fit was performed. Otherwise, 2 different linear fits were used before and after the brightest point to measure both the rise and fade rates.

Thanks to the rich optical/infrared observations available, it is tempting to use AT2017gfo as a prototypical KN to set the fading (or rising) thresholds to photometrically select KN candidates based on their photometric evolution. However, AT2017gfo is unlikely to be representative for its class, as deep follow-up and joint afterglow and KN fits of short GRBs have demonstrated \citep{Gompertz2018,AsCo2018,Rossi2020}. Instead, we chose thresholds for the fading rate based on a grid of KN models constructed with the Monte Carlo radiative transfer code \texttt{POSSIS} \citep{Bulla2019}. Specifically, we use a subset of the BNS grid presented in \cite{Dietrich2020} and the NSBH grid presented in Anand et al. (2020, in press). Both grids use geometries that are axially symmetric and constructed by varying three parameters: the mass ejected on dynamical timescales (``dynamical ejecta", $M_{\rm ej}^{\rm dyn}$), the mass released from the merger remnant and debris disk (``post-merger" ejecta, $M_{\rm ej}^{\rm pm}$) and the observer viewing angle ($\Theta_{\rm obs}$, with $\Theta_{\rm obs}=0^\circ$ corresponding to a face-on view of the system). KN light curves are predicted for 11 viewing angles from a polar ($\Theta_{\rm obs}=0^\circ$) to an equatorial ($\Theta_{\rm obs}=90^\circ$) orientation and for the following choices of the ejecta masses: $M_{\rm ej,BNS}^{\rm dyn}=[0.001,0.005,0.01,0.02]\,M_\odot$, \mbox{$M_{\rm ej,BNS}^{\rm pm}\in[0.01,0.13]\,M_\odot$} with a mass step of $0.02\,M_\odot$ and $M_{\rm ej,NSBH}^{\rm dyn}$ and \mbox{$M_{\rm ej,NSBH}^{\rm pm}\in[0.01,0.09]\,M_\odot$} with a mass step of $0.01\,M_\odot$. An half-opening angle of 30$^\circ$ is assumed for the lanthanide-rich dynamical component in both systems. This results in $308$ and $891$ KN light curves for the BNS and NSBH grid, respectively.  We refer the reader to \cite{Dietrich2020} and Anand et al. (2020, in press) for more details about the adopted BNS and NSBH grids. 

In Figure\,\ref{fig: threshold}, we show the distribution of fading rates calculated between light curve peak epoch and peak time +6\,days. In this work, we considered the threshold that constrains 95\% of models of the grid for each filter, i.e. 0.57, 0.39, and 0.27 mag\,day$^{-1}$ for $g$-, $r$-, and $i$-band respectively. All those threshold choices fully encompass the fading rate of AT2017gfo (see blue vertical lines in Figure\,\ref{fig: threshold}). 
Color evolution was not considered as a rejection criterion {\it a priori}, but it is discussed for the selected sources in \S\,\ref{sec: results}.

\subsection{Candidate vetting}
\label{subsec: candidate vetting}

Out of 11,202 objects found querying the ZTF alert database, 645 candidates with at least 3 alerts passed the selection criteria described in \S\ref{sec: photometry and detection criteria}. All candidates were assessed by visually inspecting detection image triplets (made of science image, reference image, and subtraction) and light curves (individual exposures and nightly stacks). Light curves showing erratic behavior or deep upper limits in forced photometry between detections were rejected. 
Sequences of tens of consecutive exposures were prone to show spurious ``ghost" sources that can mimic astrophysical fast transients. We effectively rejected those spurious sources aided by cone searches with \texttt{Kowalski} centered on the candidate's coordinates, that allowed us to flag for further inspection those candidates with several other alerts located along the readout channel direction within $20''$ from the central coordinates. This method was useful also to reject uncatalogued slow-moving asteroids (in any direction, not only along the readout channel direction) that fell in our sample.

\begin{figure}[t]
\centering
\includegraphics[width=1\columnwidth]{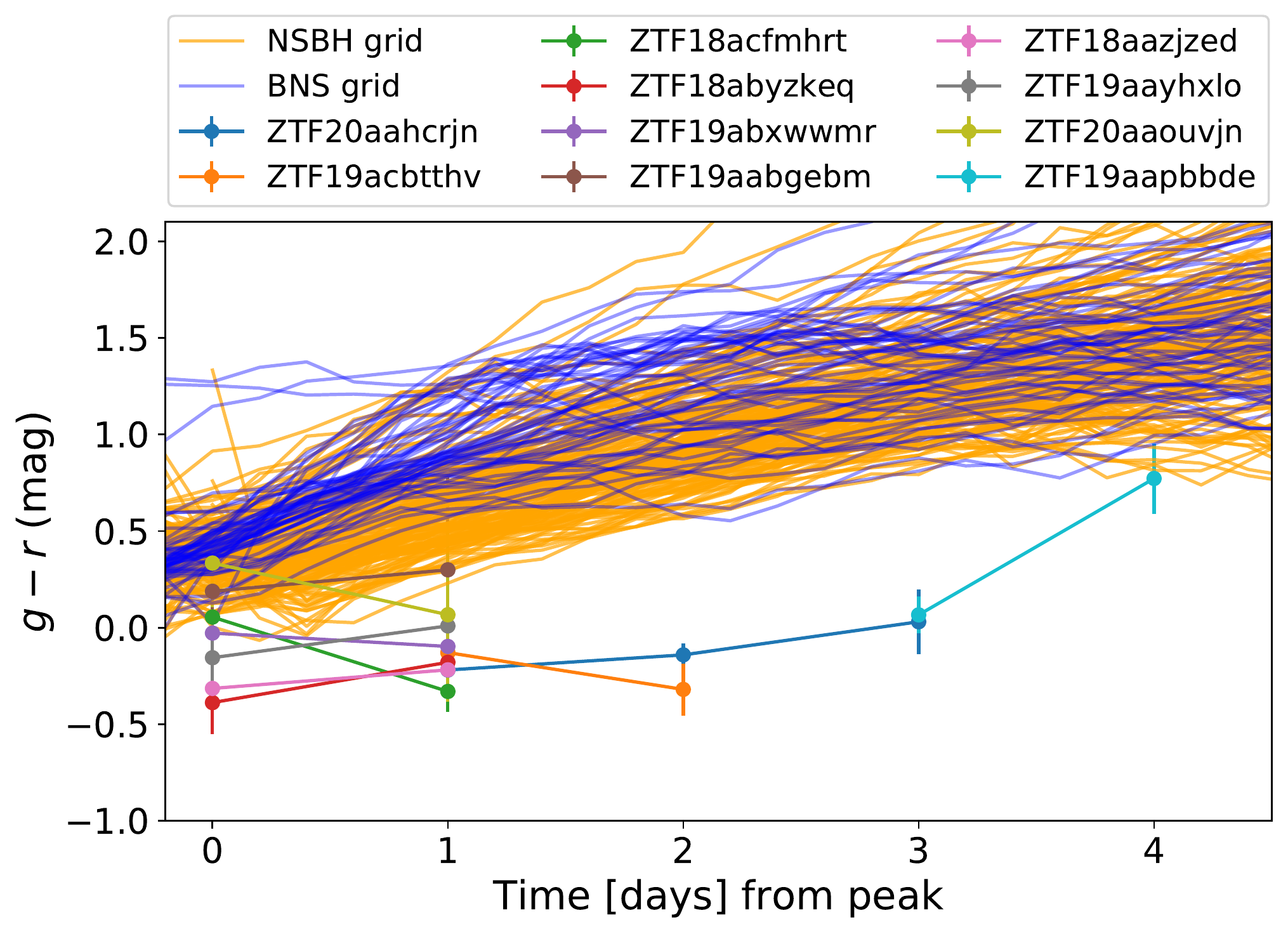}
\caption{ Here we represent the $g-r$ color evolution for BNS and NSBH grids from \cite{Bulla2019} used in this work for the candidate selection criteria. The model light curves start (time=0\,d) at $g$-band peak, while the candidates' phase is considered from the epoch of the brightest data point, irrespective of the filter. Galactic extinction effects are small in the selected ZTF fields and a correction (here not applied) would decrease the $g-r$ magnitude difference, moving the data points further from the KN models in this parameter space. The combination of color and color evolution of ZTF19aabgebm, a GRB optical afterglow, could be consistent with NSBH KN models. The behaviour of other candidates for which $g-r$ information was available on multiple nights in the decline phase appears to be distant from the KN models considered in this work in this phase space.}
\label{fig: gr_models_and_points}
\end{figure}

\section{Results}
\label{sec: results}

In this section, we present the results of our searches, divided into an all-sky search and a galaxy-targeted search. 
A total of 24 sources survived the selection criteria and strict quality checks, but none of them was deemed a robust KN candidate\footnote{Complete forced PSF photometry of these 24 sources from one week before the first detection until three weeks after the last detection can be found at \url{https://www.astro.caltech.edu/~ia/lc_candidates_IA2020_ZTF_kn_rate.csv}}. The main properties of the selected candidates are presented in Table\,\ref{table: candidates}, where candidates are grouped by i) possible crossmatch with CLU galaxies, ii) proximity to an extended source further than $1.5''$ and no underlying source, iii) presence of an underlying source within $1.5''$, and iv) ``hostless'' candidates. Appendix\,\ref{appendix: evolution rates} presents the candidates' decay rates in each band and a detailed {\it post-mortem} of individual candidates can be found in Appendix\,\ref{appendix: post-mortem}.

\subsection{All-sky search}
\label{subsec: all-sky}

The untargeted search in 514 fields, requiring $\geq 3$ detections, yielded 10,419 candidates, 23 of which passed our thorough vetting. 
Of these, 4 are spatially consistent with catalogued variable sources, 10 show variability in archival Pan-STARRS 1 \citep[PS1;][]{Chambers2016arXiv} data or in additional ZTF data, 3 are likely stellar in origin, and 3 were rejected mainly because of the combination of color and color evolution (Figure\,\ref{fig: gr_models_and_points}) being incompatible with our grid of KN models. All the candidates rejected because of their color evolution were found at relatively low Galactic latitude $|b_{\rm{gal}}| \leq 20$\,deg, which supports the hypothesis of a stellar origin. This leaves us with 3 candidates.

\begin{figure*}[t]
\centering
\includegraphics[width=0.95\columnwidth]{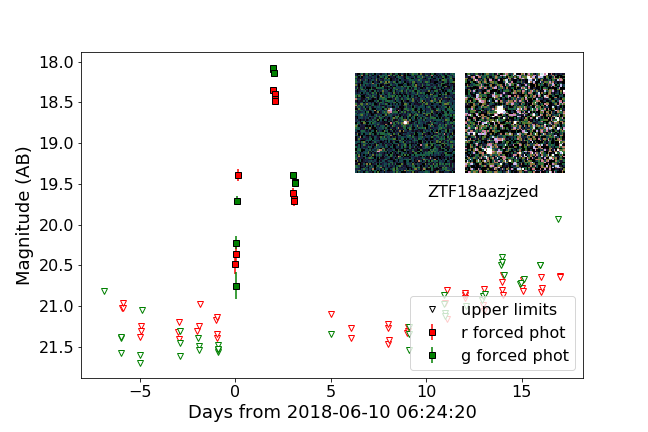}
\includegraphics[width=0.95\columnwidth]{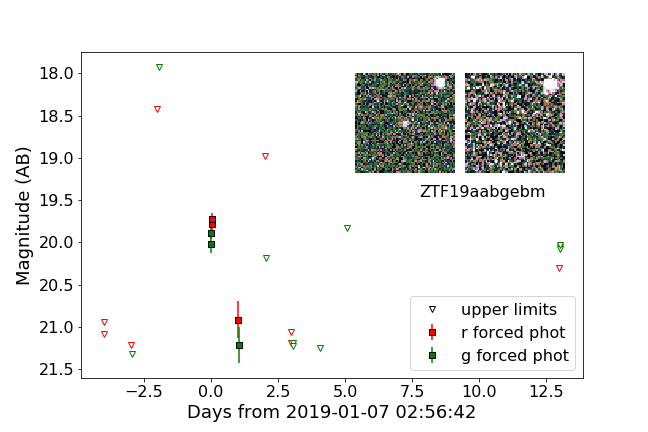}
\includegraphics[width=0.95\columnwidth]{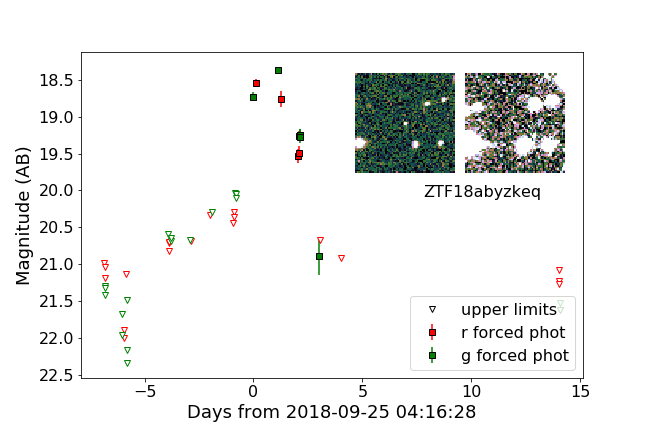}
\includegraphics[width=0.95\columnwidth]{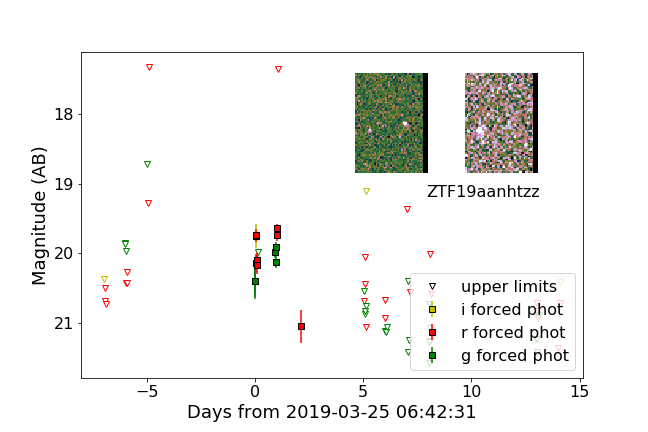}
\includegraphics[width=0.95\columnwidth]{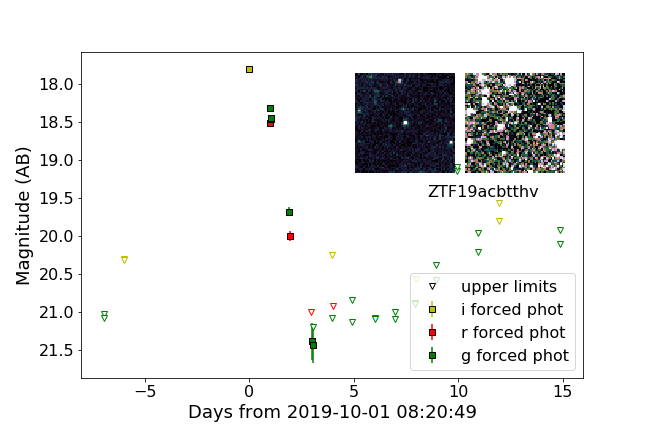}
\includegraphics[width=0.95\columnwidth]{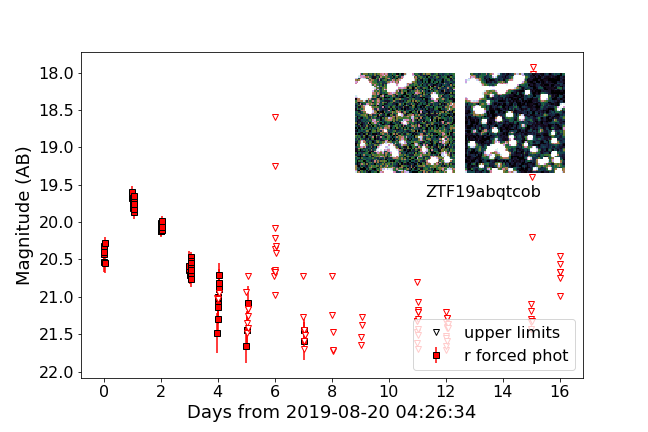}
\caption{A compilation of forced photometry light curves of candidates presented in Table\,\ref{table: candidates}. The compilation includes the GRB afterglow ZTF19aabgebm/AT2019aacx ({\it top right}) and the still mysterious fast transient ZTF19aanhtzz/AT2019aacu ({\it center right}). The inset images show the science image (left) and the reference image (right) cutouts. The cutouts are oriented with North to the top and East to the left and have a side of $60''$. }
\label{fig: light curves}
\end{figure*}

ZTF19abqtcob/AT2019aadh is located 55\,kpc (21.88$'$) from a galaxy 8.6\,Mpc away. The light curve (Figure\,\ref{fig: light curves}) shows a rapid $r$-band decay of $\alpha_r = 0.43$\,mag\,day$^{-1}$, but no color information is available.  Assuming an association with the nearby galaxy, the absolute magnitude at peak would be $M_r \sim -10$, consistent with classical nova luminosity. In addition, it is possible that ZTF19abqtcob is a foreground Galactic variable because of its low latitude ($b_{\rm gal} = -10.0$) and the lack of any other possible host galaxy visible in its proximity. Therefore, we refrain from considering ZTF19abqtcob a viable KN candidate.

The two remaining candidates, ZTF19aabgebm/ AT2019aacx and ZTF19aanhtzz/AT2019aacu, are both located at high Galactic latitude ($|b_{\rm gal}| > 36$) and show red ($g-r > 0$\,mag) color (Figure\,\ref{fig: light curves}). ZTF19aabgebm is temporally and spatially coincident with the afterglow of the long GRB~190106A \citep{Sonbas2019GCN_GRB190106A, Yurkov2019GCN_GRB190106A} at redshift $z = 1.859$ \citep{Schady2019GCN_GRB190106A}. Forced photometry allowed us to calculate a decay rate for ZTF19aabgebm of ($1.24 \pm 0.26$)\,mag\,day$^{-1}$ in both $g$ and $r$ bands. An independent identification of ZTF19aabgebm in ZTF data is described in \cite{Ho2020arXiv}.

\begin{figure}[t]
\centering
\includegraphics[width=1\columnwidth]{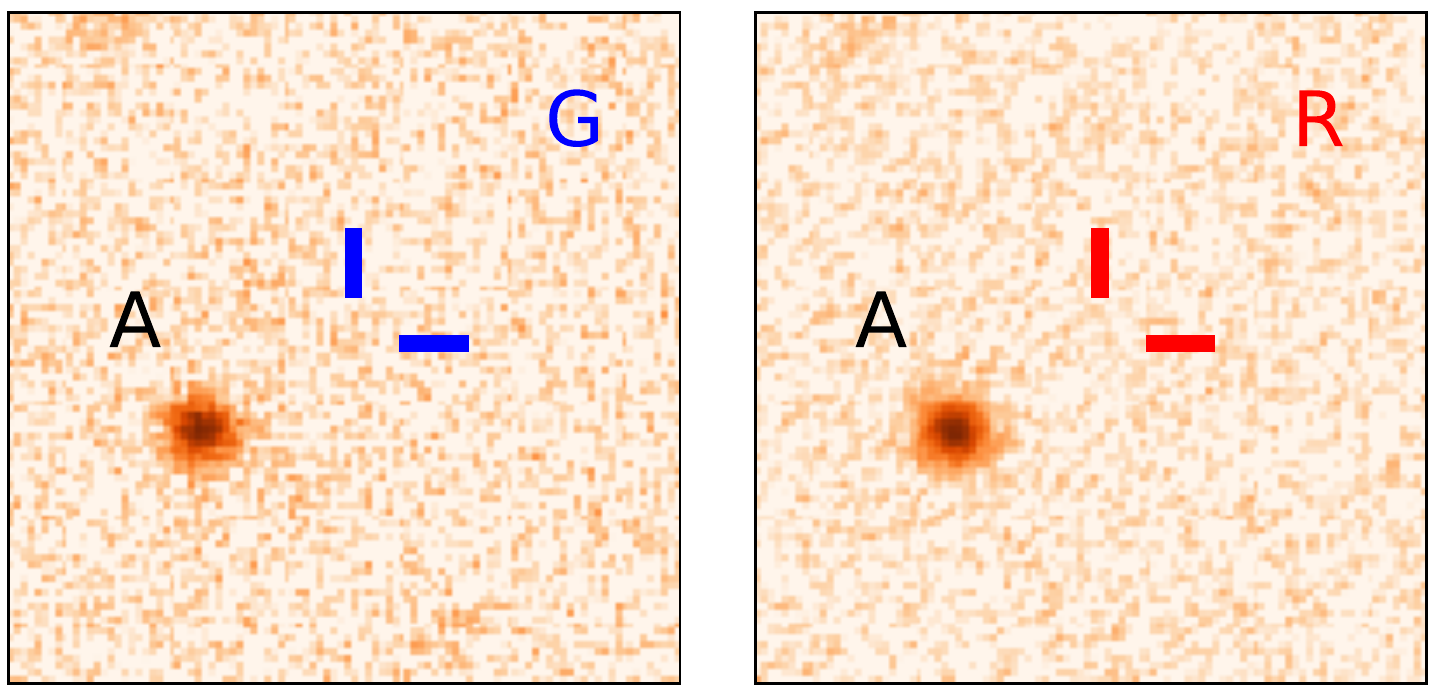}
\includegraphics[width=1\columnwidth]{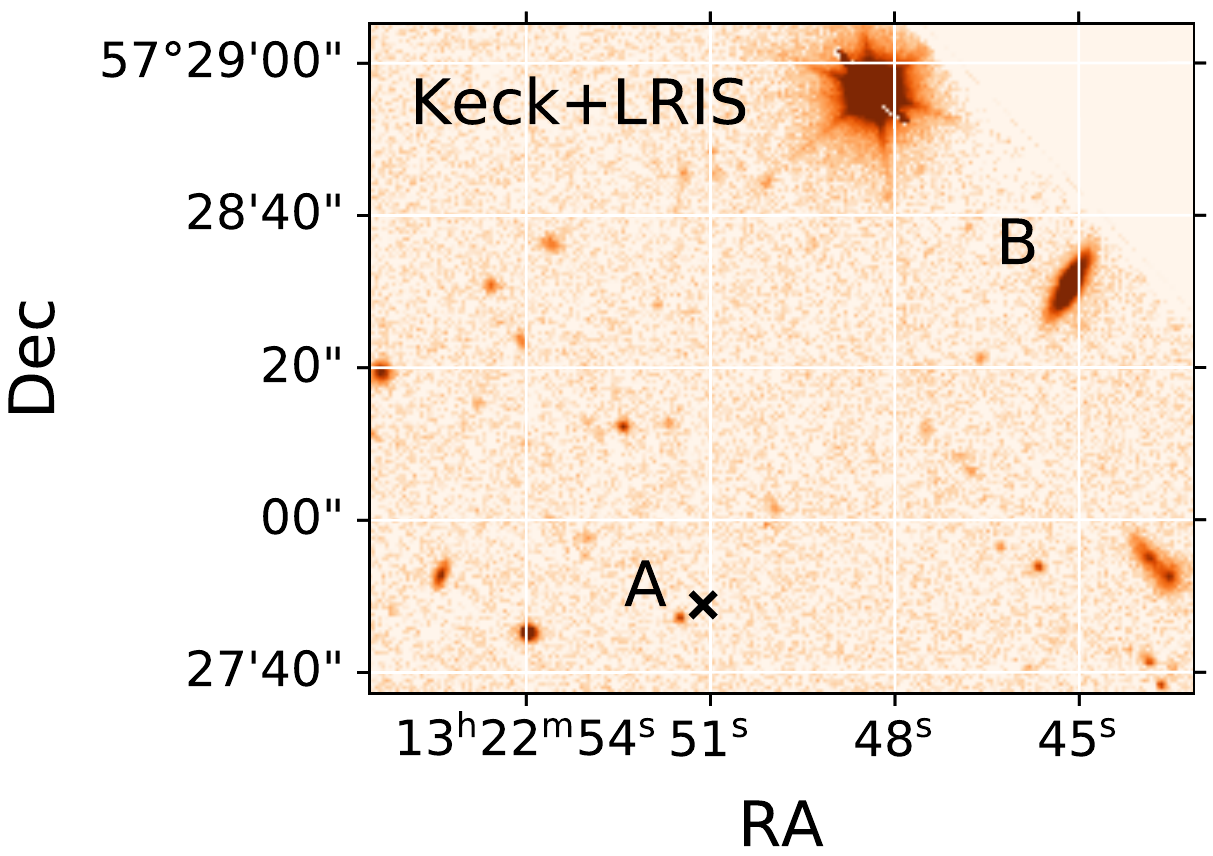}
\caption{Keck+LRIS images of the field where ZTF19aanhtzz/AT2019aacu was found. The top panels show $G$- and $R$-band images centered on the transient location, where the side of each squared image measures $13.5''$. The bottom panel shows a broader view of the field, where the transient location is marked with a black cross. All figures are oriented with North to the top and East to the left. Source ``A" is likely stellar and it is the closest object to ZTF19aanhtzz, with an angular separation of $3.5''$. Source ``B" is  WISE J132245.14+572830.4, a galaxy listed in the GLADE catalog and located at an angular separation of $1.07'$ from ZTF19aanhtzz. Our DBSP spectrum suggests the redshift of this galaxy to be $z=0.103$, which implies a projected distance of $\sim 135$\,kpc, too large for a reliable host association. }
\label{fig: keck}
\end{figure}

ZTF19aanhtzz could be considered the most puzzling transient found during our searches.  The red color ($g-r \sim 0.3$\,mag) disfavors the Galactic cataclysmic variable scenario, but there is no recorded gamma-ray signal temporally and spatially compatible with the optical transient to support the cosmological afterglow hypothesis. Deep $G$- and $R$-band images were acquired with the Low Resolution Imaging Spectrometer \citep[LRIS;][]{Oke1995} at W.~M. Keck Observatory on 2020 June 22 UT, with 600\,s of exposure time (Figure\,\ref{fig: keck}). LRIS data were processed using \texttt{lpipe} \citep{Perley2019lpipe}, a fully automatic data reduction pipeline for imaging and spectroscopy. Photometry was performed using \texttt{SExtractor} \citep{Bertin2010} and it resulted in the non-detection of a source at the transient location to 3-$\sigma$ limiting magnitudes of $G > 26.8$ and $R > 27.1$. A source is present at an angular distance of $3.5''$ in PS1 and Legacy Survey DR8 images ($r \sim 22$, labelled as source ``A" in Figure\,\ref{fig: keck}) and it can be modeled with a PSF shape according to the Legacy Survey DR8 catalog \citep{Dey2019}, which suggests the source to be stellar. No extended emission from source A is clearly detectable in the LRIS images. 

The galaxy WISE J132245.14+572830.4, labelled as source ``B" in Figure\,\ref{fig: keck}, is placed at a distance of 267\,Mpc by the Galaxy List for the Advanced Detector Era catalog \citep[GLADE;][]{Dalya2018}, which would imply a projected distance of $\sim 83$\,kpc between the galaxy and ZTF19aanhtzz, if it was indeed the transient's host. Such a distance is within the range of projected offsets for neutron star mergers, considering short GRBs as a proxy \citep{Berger2014sGRB}. 
On 2020-07-02 UT, we acquired a spectrum of WISE J132245.14+572830.4 using the Double Spectrograph (DBSP) mounted on the Palomar 200-inch telescope with 600\,s of exposure time. The data were processed with a custom \texttt{PyRAF} reduction pipeline for DBSP \citep{Bellm2016DBSP}. The spectrum shows prominent Balmer series and other common galaxy features that place the galaxy at redshift $z=0.103$, i.e. at a luminosity distance of 490.57\,Mpc and a projected distance of $\sim 135$\,kpc from the transient. The projected distance is too large for WISE J132245.14+572830.4 to be a reliable host of ZTF19aanhtzz \citep[although short GRB were found with a projected distance from the host larger than 100\,kpc;][]{Berger2014sGRB}.

If neither source A or Source B are the transient's host, we consider the hypothesis of an underlying host galaxy being too faint to be detectable in our Keck+LRIS images. If such a faint host was located at a distance of $<250$\,Mpc, its luminosity would be $M\sim -10$ or fainter, typical of dwarf galaxies satellite of more massive galaxies. The apparent lack of such a massive galaxy in the field of ZTF19aanhtzz makes this scenario improbable. It is possible that a more massive host galaxy at high redshift exits that was not detectable in our Keck+LRIS images. If ZTF19aanhtzz was a Galactic source, located in the outskirts of the Milky Way, the LRIS limit suggests the absolute magnitude of the quiescent counterpart to be $M > 7$\,mag. Based on a typical HR diagram, this only allows a main sequence star that is later than K0 in spectral type, or a white dwarf, to be the progenitor star, but a M-dwarf progenitor is likely excluded by the non-detection in $z$-band by the Legacy Survey.
We consider now the scenario in which ZTF19aanhtzz was a cosmological relativistic afterglow. The lack of a detected GRB and the observation of the rise phase of the transient make it unlikely for ZTF19aanhtzz to be the afterglow of a relativistic explosion viewed on-axis. Although a detail analysis using off-axis GRB models \citep[e.g.,][]{VanEerten2010, Beniamini2020off} is outside the scope of this paper, we suggest that a slightly off-axis relativistic explosion might explain the behaviour of this transient.
In summary, with the data in hand, a conclusive answer regarding the nature of ZTF19aanhtzz/AT2019aacu is yet to be found, but a cosmological afterglow origin scenario is favored.

\subsection{Galaxy-targeted search}
\label{subsec: galaxy-targeted}

In addition to all-sky searches, we conducted deeper galaxy-targeted queries by requiring at least two detections (instead of three detections) to be present for each candidate. We found that 92 candidates passed the automatic selection criteria and could be spatially cross-matched with CLU galaxies $>10$\,Mpc away within a 100\,kpc radius. Of those, 3 candidates pass the visual inspection and quality checks, namely ZTF19abzwbxy, ZTF20aaouvjn, and ZTF18achqdzh (Table\,\ref{table: candidates}). The former two were also found during the all-sky search (see \S\ref{subsec: all-sky}), while this galaxy-targeted search yielded the new candidate ZTF18achqdzh, a bright transient ($M_r < -17$) likely generated by nuclear activity. All these candidates were ruled out. 
We conclude that none of our candidates can be considered a viable KN associated with nearby ($<200$\,Mpc) cataloged galaxies.

\section{Discussion}
\label{sec: discussion}

In this work, we used rigorous criteria to identify uncommon transients that could possibly be KNe. The rich ZTF dataset provided us with high cadence photometry and, usually, color information that we could use to find rapidly fading candidates possibly associated with nearby galaxies and separate them from other classes of transients or variables. Our searches did not yield any source that we claim to be a viable KN candidate, although the nature of some fast transients remains unknown, including the possibly extragalactic fast transient ZTF19aanhtzz. It is possible that KNe were present in the dataset, but they were missed in this analysis because, for example, they were too close to the detection limit, due to the pipeline detection efficiency, or because they evolved in different ways than we expected based on available models and on the knowledge of GW170817.

This search for KNe in the ZTF survey, independent of GW or short GRB triggers, complements the ZTF searches for optical counterparts to 13 neutron start mergers found in GWs during O3, described in \cite{Kasliwal2020arXiv}.  In that work, we used the query and photometry methods described in \S\ref{sec: methods} to search for KNe by requiring $\geq 1$ detection in the alert stream over the 95\% of the localization skymaps in the 3 days following the mergers. The analysis presented in \cite{Kasliwal2020arXiv} led to no viable KN counterpart to O3 neutron star mergers.

 This work further demonstrated the importance of repeated, deep observations of wide-field surveys to better understand the nature of newly identified astronomical transients. A key problem in KN searches in ``archival" survey data is the ability to recognize KNe without spectroscopic follow-up. The problem of photometric identification of KNe among other transients was already addressed, for example, by \cite{Cowperthwaite2015, Doctor2017, Cowperthwaite2018, Bianco2019PC}. In general, these works find that a combination of color information (at a given time) and rapid evolution is effective at separating KNe from other extragalactic sources such as supernovae. In our case, the selection criteria on the fading rate and transient duration had already helped us remove all supernova candidate from our sample, leaving AGN and Galactic sources as main contaminants, as expected from previous fast transient searches \citep[e.g.,][]{vanRoestel2019, Andreoni2020MNRAS}. 
 
 We use the grid of models described in \S\ref{sec: photometry and detection criteria} to visualize the expected $g-r$ color evolution over time for BNS and NSBH mergers. Figure\,\ref{fig: gr_models_and_points} shows that all those candidates for which inter-night color evolution is available have color evolution different than predicted by KN models. Interestingly, the candidate that approaches KN models the most is ZTF19aabgebm, the red and rapidly-evolving afterglow of GRB~190106A.
We also considered a set of 30 KN models from the grid computed by \cite{Kasen2017Nat} with parameters similar to the blue component of AT2017gfo in the ranges $M_{\rm ej}\in[0.01,0.04]\,M_\odot$, $v_k \in [0.1-0.3]$, and $X_{\rm lan} \in [1{\rm e}-5, 1{\rm e}-4]$ for ejecta mass, kinetic ejecta velocity, and lanthanide fraction, respectively. 
In the first 3 days from the merger, these models are characterized by blue colors and thus provide better match to our KN candidates compared to the \texttt{POSSIS} grid shown in Figure\,\ref{fig: gr_models_and_points}. However, the resulting light curves get redder even more steeply than the \texttt{POSSIS} grid, placing the candidates even further from the region of the phase space expected for KNe at later epochs.

In general, systematic searches in the broadest possible parameter space covering the models presented in Section~\ref{sec: photometry and detection criteria} yield reasonable fits to our selected candidates \citep{CoDi2018,StCo2019}; the lack of either explosion time or distance constraints make it possible to fit them within error bars of the models. This emphasizes the need for rapid photometric follow-up of sources of this type to make the most straightforward classifications.

\begin{table*}
    \centering
    \begin{tabular}{lrrrrccl}
    \hline \hline
Name & RA & Dec & $b_{\text{gal}}$ & distnr & sgscore & CLU match & Comment \\
Name & (deg) & (deg) & (deg) & (arcsec) &  & (boolean) &  \\
\hline
ZTF18aazjzed\footnote{Within 100\,kpc from a galaxy $<10$\,Mpc away} & 245.80994 & 65.01023 & 39.8 & - & - & 1 & CV \\
ZTF19abzwbxy\footnote{Found in both all-sky and galaxy-targeted searches} & 305.86048 & 6.66687 & $-17.0$ & $<1.5$ & 0.13 & 1 & QSO \\
ZTF18achqdzh\footnote{Found only in galaxy-targeted search} & 40.20850 & $-9.17101$ & $-58.5$ & $<1.5$ & 0.12 & 1 & Bright nuclear variability \\
ZTF20aaouvjn$^{\rm{b}}$ & 169.87812 & 15.41104 & 65.7 & - & - & 1 & Multiple outbursts in ZTF \\
ZTF19abqtcob$^{\rm{a}}$ & 295.23436 & 2.22770 & $-10.0$ & - & - & 1 & Extragalactic nova or foreground CV? \\
\hline
ZTF18abyzkeq & 342.39259 & 37.93156 & $-19.0$ & 14.35 & 0.10 & 0 & CV \\
ZTF19aayhxlo & 295.77816 & 48.32182 & 12.0 & 8.85 & 0.06 & 0 & Post-detections in ZTF \\
ZTF19acecsfi & 119.87837 & 44.27264 & 30.3 & 23.62 & 0.02 & 0 & Likely stellar/CV  \\
ZTF19aabgebm & 29.88002 & 23.84546 & $-36.4$ & 9.09 & 0.01 & 0 & GRB190106A afterglow  \\
ZTF20aahcrjn & 109.06754 & 35.75254 & 20.2 & 9.70 & 0.03 & 0 & Multiple bursts in ZTF \\
\hline
ZTF18abuzpri & 288.24987 & 61.55687 & 21.2 & $<1.5$ & 0.32 & 0 & Stellar \\
ZTF19abqiwjq & 293.06491 & 6.72184 & $-5.9$ & $<1.5$ & 0.38 & 0 & Pre-activity in PS1 images\\
ZTF19abqneae & 295.60303 & 0.56979 & $-11.1$ &$<1.5$ & 0.00 & 0 & Pre-activity in PS1 images\\
ZTF19acbtthv & 16.02320 & 50.12632 & $-12.7$ & $<1.5$ & 0.49 & 0 & Pre- and post-detections in PS1 and ZTF\\
ZTF19acszwgx & 271.89500 & 10.59102 & 14.5 & $<1.5$ & 0.03 & 0 & Pre-activity in PS1 images\\
ZTF19abcputm & 294.13549 & $-7.35038$ & $-13.4$ & $<1.5$ & 0.14 & 0 & Pre-activity in PS1 images\\
ZTF19abogdfr & 293.39693 & $-17.64588$ & $-17.1$ & $<1.5$ & 0.08 & 0 & Pre-activity in PS1 images\\
ZTF19aapbbde & 270.30443 & 13.28065 & 17.0 & $<1.5$ & 0.37 & 0 &  Pre-activity in PS1 images \\
\hline
ZTF19abcpiag & 264.64519 & 1.42066 & 16.8 & - & - & 0 & Catalogued variable source \\
ZTF18acfmhrt & 30.63920 & 55.41712 & $-6.1$ & - & - & 0 & Color evolution unlike KN \\
ZTF19abxwwmr & 313.90428 & 22.98274 & $-14.2$ & - & - & 0 & Color evolution unlike KN\\
ZTF19aanhtzz & 200.71300 & 57.46357 & 59.2 & - & - & 0 & Cosmological afterglow? \\
ZTF19abudvoz & 46.68777 & 46.15358 & $-10.6$ & - & - & 0 & Pre-activity in PS1 images\\
ZTF18acsjqjd & 29.91694 & 54.17361 & $-7.4$ & - & - & 0 &  Pre-activity in PS1 images \\
    \hline
    \end{tabular}
    \caption{Candidates that passed our selection criteria described in \S\ref{sec: photometry and detection criteria}. Along with the equatorial J2000 coordinates and Galactic latitude ($b_{\rm{gal}}$), we indicate the distance to the closest source within 30$''$ with star/galaxy classification score $\rm{sgscore} \leq 0.1$, suggesting the source to be a possible galaxy. The star/galaxy classification score of those candidates lying on top of faint sources (${\rm distnr} < 1.5''$) is reported even where $\rm{sgscore} > 0.1$. The 24 candidates are grouped by their possible crossmatch with CLU galaxies, proximity to an extended source further than $1.5''$ and no underlying source,  presence of an underlying source within $1.5''$, and finishing with ``hostless'' candidates. Finally, the comment column summarizes the {\it post-mortem} description of each candidate from \S\ref{appendix: post-mortem}.}
    \label{table: candidates}
\end{table*}

\section{Kilonova rates}
\label{sec: rates}

The searches conducted during the first 23 months of the ZTF survey, based off of alerts, have not yielded any likely KN candidate. This result can be used to constrain KN rates to compare against neutron star merger rates obtained from GW and short GRB observations.

Rates were estimated using \texttt{simsurvey} \footnote{\url{https://simsurvey.readthedocs.io/en/latest/}} \citep{Feindt2019}, a simulation software package that accounts for custom transient light curves and observational parameters of the survey. 
\texttt{simsurvey} uses Monte Carlo methods to simulate synthetic lightcurves. We used ZTF data to calculate the number of KNe detected for each rate used in the injection of synthetic transients. Milky-Way extinction was applied based on the \cite{Schlegel1998} reddening maps. 
We considered three types of KN light curves: a 3-day constant luminosity (or ``top-hat") model, a linear decay model, and models based on radiative transport simulations. Synthetic sources are injected in \texttt{simsurvey} assuming a broad range of rates. Our limits on the rate are chosen at the point where three of the injected KNe are recovered, as dictated by Poissonian small number statistics for a 95\% confidence level \citep{Gehrels1986}. 

\begin{figure*}[t]
\centering
\includegraphics[width=\textwidth]{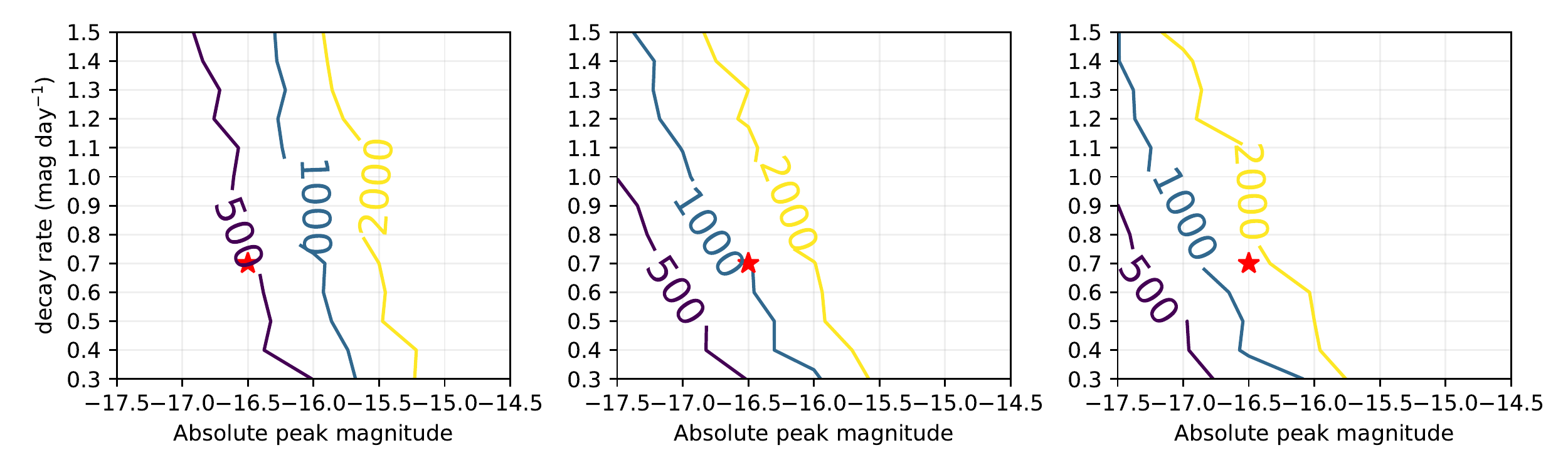}
\caption{Estimated upper limits on the KN rate (Gpc$^{-3}$\,yr$^{-1}$, indicated by colored contours) per peak absolute magnitude and linear decay rate requiring 1 ({\it left}), 2 ({\it middle}) and 3 ({\it right}) detections ($5\sigma$) for each light curve at a 95\% confidence level. The red star marker indicates where a blue KN similar to AT2017gfo would lie if it had a starting absolute magnitude of $M=-16.5$ and a linear decay rate of 0.7\,mag\,day$^{-1}$, which is the approximate average between the measured decay rates in $g$- and $r$-bands. As expected, the slower/brighter KNe are, the more stringent the limit on rates from ZTF is.}
\label{fig:rate3}
\end{figure*}

\begin{figure*}[t]
\centering
\includegraphics[width=\textwidth]{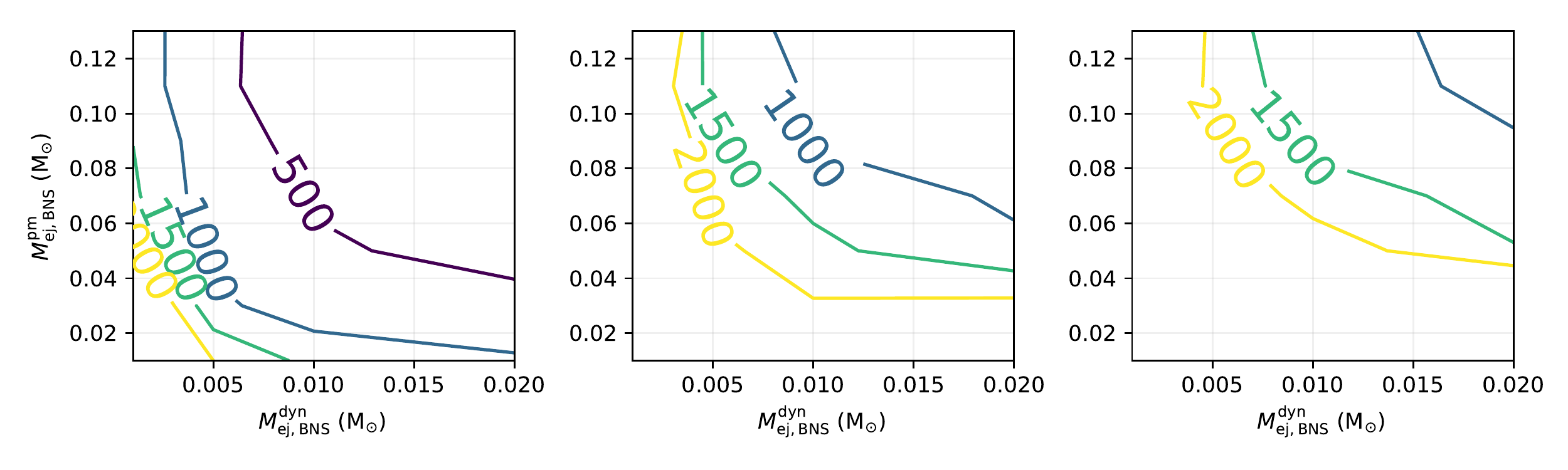}
\includegraphics[width=\textwidth]{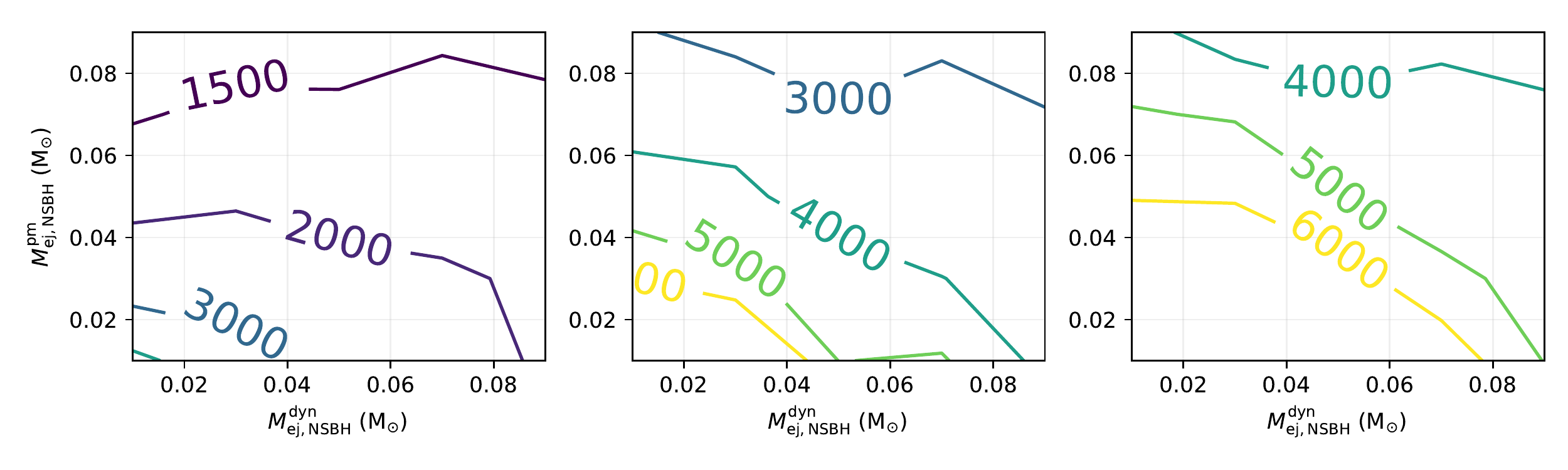}
\caption{Estimated upper limits on the KN rate (Gpc$^{-3}$\,yr$^{-1}$, indicated by colored contours) per ejecta mass component from BNS ({\it top}) and NSBH ({\it bottom}) grids requiring 1 ({\it left}), 2 ({\it middle}) and 3 ({\it right}) detections ($5\sigma$) for each light curve at a 95\% confidence level.}
\label{fig: mej grid}
\end{figure*}

\begin{itemize}
    \item {\it Top-hat --} Constant 3-day light curve model (Appendix\,\ref{appendix: rates grid tophat}). The top-hat model at $M_{g,r,i}=-16$ (about the maximum brightness measured for AT2017gfo) yielded a rate of  $R<398$\,Gpc$^{-3}$\,yr$^{-1}$ for at least one $5\sigma$ detection, $R<930$\,Gpc$^{-3}$\,yr$^{-1}$ for at least two $5\sigma$ detections separated by at least 3\,hr, and $R<1121$\,Gpc$^{-3}$\,yr$^{-1}$ by requiring at least three $5\sigma$ detections.  Appendix\,\ref{appendix: rates grid tophat} presents a table with the results for a range of absolute magnitudes $M \in [-12, -17]$.   

    \item {\it Linear decay --} We considered a grid of linear decay models by varying the starting absolute magnitude $M \in [-14.5, -17.5]$\,mag and the decay rate $\alpha \in [0.3, 1.5]$\,mag\,day$^{-1}$ to encompass the most probable decay rates that we considered to chose our selection criteria (Figure\,\ref{fig: threshold}, Appendix\,\ref{appendix: rates grid linear}). 
    These models do not account for color evolution and are filter-agnostic. The table in Appendix\,\ref{appendix: rates grid linear} presents the results for the linear decay grid and
    Figure\,\ref{fig:rate3} represents contours indicating rates of $R = 500$, 1000, and 2000\,Gpc$^{-3}$\,yr$^{-1}$.  
    
    \item {\it Radiative transport simulations --} We use the best-fit model for AT2017gfo obtained in \cite{Dietrich2020} and assume two KN populations: the first where all KNe have light curves like AT2017gfo, the second where KNe are instrincally like AT2017gfo, but with viewing angle $i$ uniformly distributed in cos($i$). The results are shown in Appendix\,\ref{appendix: rates AT2017gfo}. In addition, we explore the parameters space of ejecta masses for the models tailored to BNS and NSBH from \cite{Dietrich2020} and Anand et al. (2020, in press), also described in \S\ref{sec: photometry and detection criteria}. The resulting rates for the grid of models obtained with a fixed half-opening angle $\phi = 30$\,deg are shown in Appendix\,\ref{appendix: rates mej grid} and illustrated in Figure\,\ref{fig: mej grid}. We note that the NSBH grid provides less constraining results. The light curves are fainter at peak, and broader compare to the BNS grid. Therefore, the detected light curves would be observable for longer times, but in a smaller distance range.

\end{itemize}

\begin{figure*}[t]
\centering
\includegraphics[width=\textwidth]{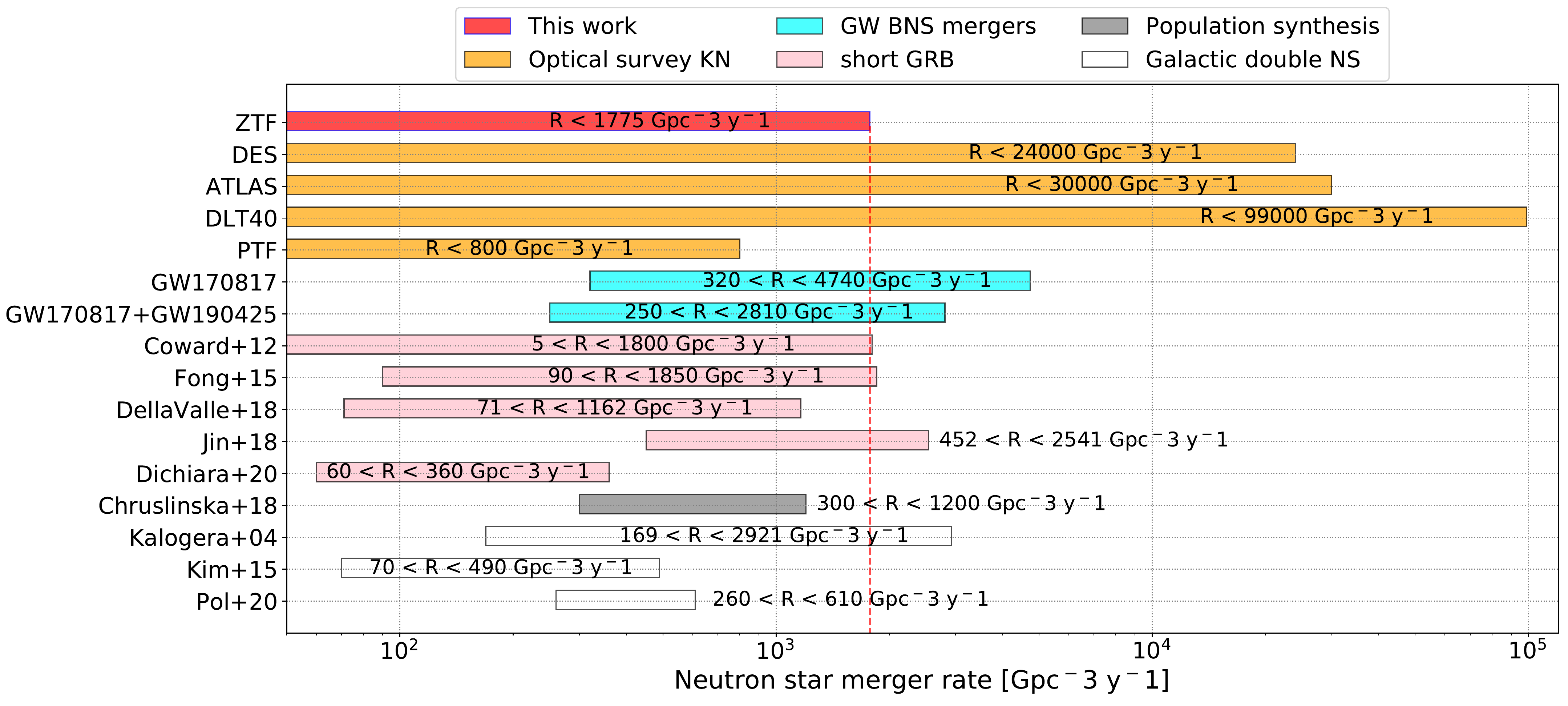}
\caption{The constraints on the KN rate obtained in this work (red bar, dashed red line) can be compared with the results of other projects \citep{Kasliwal2017, Smartt2017, Yang2017} that considered a uniform KN population similar to GW170817 \citep[barred the DES result, produced before the discovery of GW170817;][]{Doctor2017}. If all BNS mergers are accompanied by KNe as bright as GW170817, then our results could improve the constraints on the BNS merger rate obtained from GW observations \citep{Abbott2017GW170817discovery, Abbott2020GW190425discovery}. In fact, the true KN luminosity function is still poorly understood. The plot also includes some examples of neutron star merger rates from short GRBs \citep{Coward2012, Fong2015, DellaValle2018, Jin2018, Dichiara2020}, population synthesis \citep{Chruslinska2018}, and BNS observations in the Milky Way  \citep{Kalogera2004erratum, Kim2015, Pol2020} for comparison.}
\label{fig: rates comparison}
\end{figure*}

 The rates obtained by requiring three $5\sigma$ detections represent our all-sky search conservatively. In fact, we queried the ZTF database selecting sources with at least 3 detections in the alerts, however only one alert (chronologically the last one) needs to have $\geq 5\sigma$ significance. For example, it is possible that only 1 alert is issued for a transient when its flux passes the $5\sigma$ threshold, but the alert packet includes information about low-significance detections (between $3\sigma$ and $5\sigma$) that did not trigger any alert generation. These low-significance detections were accounted for during our query. In addition, our experiment targeting CLU galaxies required only 2 detections at least 3\,hr apart, the first of which could be a low-significant detection below the $5\sigma$ mark.
 
 On the other hand, \texttt{simsurvey} does not account for the detection efficiency curve of the ZTF data processing pipeline, which is currently unavailable. \cite{Frohmaier2017} calculated the detection efficiencies for the Palomar Transient Factory (PTF) pipeline \citep[that relies on the image-subtraction algorithm developed by][]{Alard1998}, obtaining a completeness of $\sim 95\%$ where the host galaxy surface brightness is $\lesssim 10\%$ of the transient brightness. From the PTF experience, we can also expect the detection completeness to drop to $60\%-70\%$ in the presence of bright hosts with $r \lesssim 18$\,mag \citep{Frohmaier2017}. However, a dedicated study of the recovery efficiency of the ZTF pipeline \citep[that relies on the \texttt{ZOGY} image-subtraction algorithm;][]{Zackay2016} is necessary to provide exact completeness figures. Such a study lies outside the scope of this paper. 
 
  We estimated that the random misalignment between science and reference images can cause a loss of $\sim 1\%$ of the chip area. Also, \cite{Fremling2020RCF} calculated
that < 1\% of astronomical transients will be lost
due to proximity to bright sources outside the Galactic plane.
 
 Catalog incompleteness should also be taken into account. A galaxy number completeness lower limit of $57\%$ integrated between $0 < z < 0.05$ was obtained for the NASA/IPAC Extragalactic Database (NED) by \cite{Fremling2020RCF} using Type Ia supernovae as proxy. The CLU catalog encompasses NED and includes $>18,000$ galaxies within 200\,Mpc found during the PTF H$\alpha$ survey \citep{Cook2017CLU}, so we expect our integrated completeness to be larger than 57\%. 
 
 With these caveats in mind, we will now discuss the main results obtained in this analysis, considering the case of 2 $5\sigma$ detections as the best representation of our results.

 Using the best-fit model to AT2017gfo in radiative transfer simulations, for two $5\sigma$ detections, we obtained a rate of $R < 1775$\,Gpc$^{-3}$\,yr$^{-1}$, while considering a population of KNe with the same intrinsic properties of AT2017gfo and with a uniform viewing angle distribution, we obtained a rate of $R < 4029$\,Gpc$^{-3}$\,yr$^{-1}$.
 As discussed in \S\ref{sec: results}, models obtained with \texttt{POSSIS} radiative transport simulations tend to err towards redder and fainter KNe. Assuming a blue KN with a linear decay rate of $\alpha = 0.7$\,mag\,day$^{-1}$ as for AT2017gfo between $g$- and $r$-band with a peak magnitude of $M = -16.5$, we can constrain the KN rate to be $R < 937$\,Gpc$^{-3}$\,yr$^{-1}$ (1363\,Gpc$^{-3}$\,yr$^{-1}$) by requiring 2 (3) detections in ZTF data ($5\sigma$). 

Before the discovery of GW170817, \cite{Doctor2017} constrained the bright KN rate to be $R <  2.4 \times 10^4$\,Gpc$^{-3}$\,yr$^{-1}$ (90\% confidence) with the Dark Energy Survey. Then,
other surveys constrained the KN rate using AT2017gfo as reference.
\cite{Smartt2017} set the KN rate to be $R < 3.0 \times 10^4$\,Gpc$^{-3}$\,yr$^{-1}$ (95\% confidence) with ATLAS and \cite{Yang2017} placed an upper limit of $R < 9.9 \times 10^{4}$\,Gpc$^{-3}$\,yr$^{-1}$ (90\% confidence) with 13 months of nightly monitoring of nearby galaxies with DLT40. The limits calculated in this work are more constraining than those listed above. A more stringent limit of $R < 800$\,Gpc$^{-3}$\,yr$^{-1}$ was placed by \cite{Kasliwal2017} with PTF data, also focusing on galaxies present in the CLU catalog (although a correction for the galaxy catalog completeness was not applied).
Our results are consistent with ZTF survey simulation predictions described in \cite{SaguesCarracedo2020arXiv}. The non-detection of a KN in almost 2 years of survey suggests that the number of KNe detectable with ZTF estimated by \cite{Scolnic2018} (using a fixed rate of $R = 1000$\,Gpc$^{-3}$\,yr$^{-1}$ and average values for the cadence, zero point, and sky noise) may have been too optimistic.

When comparing KN rates with BNS merger or short GRB rates, a KN luminosity function should be considered, but its uncertainties are still large \citep{Kasliwal2020arXiv}. 
Our limits using a GW170817-like model are consistent with isotropic equivalent short GRB rates including, for example,
$R = 270^{+1580}_{-180}$\,Gpc$^{-3}$\,yr$^{-1}$ \citep{Fong2015},
$R = 352^{+810}_{-281}$\,Gpc$^{-3}$\,yr$^{-1}$ \citep{DellaValle2018},
or $R = 160^{+200}_{-100}$\,Gpc$^{-3}$\,yr$^{-1}$ \citep{Dichiara2020}.
ZTF results are also consistent with rates estimated with population synthesis models such as $R = 600^{+600}_{-300}$\,Gpc$^{-3}$\,yr$^{-1}$ \citep{Chruslinska2018}, a solution among those that returned higher rates that are consistent with the BNS merger rate from GWs \citep{Abbott2017GW170104discovery, Abbott2020GW190425discovery}. 

Assuming that all BNS mergers generate a bright optical transient, the ZTF limits can constrain the upper side of rates obtained via GW observations of 250--2810\,Gpc$^{-3}$\,yr$^{-1}$ \citep{Abbott2020GW190425discovery}, under the assumption that most KNe are intrinsically similar to AT2017gfo. The ZTF KN rate can also constrain the rates of $R = 1109^{+1432}_{-657}$\,Gpc$^{-3}$\,yr$^{-1}$ \citep{Jin2018}, calculated from short GRB observations, and $R = 830^{+2091}_{-661}$\,Gpc$^{-3}$\,yr$^{-1}$, derived from BNS system observations in the Milky Way 
\citep[$R_{\rm{BNS}} = 83.0^{+209.1}_{-66.1}$\,Myr$^{-1}$ at $95\%$ confidence level;][]{Kalogera2004erratum} assuming a galaxy number density of $\sim 10^{-2}$\,Mpc$^{-3}$ as in \cite{DellaValle2018}. Our results are consistent with BNS merger rates obtained by \citealt{Kim2015}, who calculated $R_{\rm{BNS}} = 21^{+28}_{-14}$\,Myr$^{-1}$ at $95\%$ confidence level. Similarly, our result is compatible with the updated Milky Way BNS merger rate obtained by \citealt{Pol2019, Pol2020} of $R_{\rm{BNS}} = 37^{+24}_{-11}$\,Myr$^{-1}$ at $90\%$ confidence level. The rates from \citealt{Kim2015} and \citealt{Pol2020} 
translate into $R = 210^{+280}_{-140}$\,Gpc$^{-3}$\,yr$^{-1}$ and $R = 370^{+240}_{-110}$\,Gpc$^{-3}$\,yr$^{-1}$, respectively, using again a galaxy number density of $10^{-2}$\,Mpc$^{-3}$. 

Figure\,\ref{fig: rates comparison} offers a visual representation\footnote{A ``living" version of Figure\,\ref{fig: rates comparison} can be found at \url{https://www.astro.caltech.edu/~ia/plot_gantt_chart_rates.pdf}} of the rates discussed in this section. 

If changes in the viewing angle make the KN rate become $R < 4029$\,Gpc$^{-3}$\,yr$^{-1}$ as we predicted using radiative transfer models \citep{Bulla2019}, our ZTF limits would become unconstraining for BNS mergers. The current rate from GWs does not account for NSBH mergers, that also are expected to be accompanied by bright KNe under favorable conditions of mass ratios and neutron star radii, thus it could be interesting to compare the ZTF constrains with new BNS + NSBH merger rates from GW observations in the near future.

\section{Conclusions}
\label{sec: conclusion}

We explored 23 months of ZTF survey data searching for rapidly-fading transients that could be classified as KNe based on their photometric evolution. In addition to ZTF alert database queries, we employed forced PSF photometry and nightly flux stacking to improve our ability of recognizing KNe among $>10^4$ candidates. All most promising transients turned out to be either Galactic foreground or cosmological background sources, leaving no viable KN in the sample. The observations allowed us to place constraints on the KN rate for a variety of radiative transport based models, specifically $R < 1775$\,Gpc$^{-3}$\,yr$^{-1}$ for KNe similar to AT2017gfo and $R < 4029$\,Gpc$^{-3}$\,yr$^{-1}$ for KNe with the same intrinsic properties of AT2017gfo assuming a uniform viewing angle distribution.

Several on-going surveys observe large volumes of sky at increasingly high cadence including, for example, Pan-STARRS \citep{Chambers2016arXiv}, the Asteroid Terrestrial-impact Last Alert System (ATLAS\footnote{\url{http://atlas.fallingstar.com/}}), and All-Sky Automated Survey for Supernovae \citep[ASAS-SN;][]{Shappee2014, Holoien2017}, ZTF itself, and multi-facility programs such as ``Deeper, Wider, Faster" \citep[][Cooke et al. in preparation;]{Andreoni2020MNRAS}. 
Dedicated searches in these large datasets may be necessary to find KNe \citep[see for example][regarding on-going KN searches with PS1]{Smartt2019TNSAN, McBrien2020arXiv}. 

Based on this study, it is clear that both increased depth and rapid follow-up of interesting candidates, as executed for example in the ZTF follow-up of gravitational-wave events \citep[][and Anand et al., in press]{Coughlin2019GW190425, Kasliwal2020arXiv}, should be prioritized to maximize the probability of a KN discovery. In about 2 years, ZTF has allowed us to place tighter constraints on the KN rate than most other surveys, but there is room for improvement in the near future. Two more years of ZTF survey will naturally halve the rate limits that we calculated. To maximize the volume explored, depth should be prioritized over area \citep[e.g.,][]{Ghosh2017}. Nightly stacking of ZTF images (Goldstein et al., in preparation) or longer exposures may be successful avenues for KN discovery.  The identification of KN candidates in real time is indisputably valuable because it can trigger rapid multi-wavelength follow-up observations, including optical/IR spectroscopy. On the other hand, the limited availability of 8-m class telescopes, necessary to classify sources fainter than $\sim 21$\,mag, represents a challenge for discovery programs relying on deeper observations. 

This work has shown that photometric monitoring combined with archival information from multiple surveys can also be very effective at separating promising KN candidates from other types of rapidly evolving transients. The ability to recognize KNe solely based on their light curves will be key during the upcoming Legacy Survey of Space and Time \citep[LSST;][]{Ivezic2019LSST} at the Vera C. Rubin Observatory. Photometric KN candidate identification will be possible, in particular, if a ``rolling" cadence is chosen that entails nightly multi-band imaging \citep{Andreoni2019LSST}. Although on-going surveys have some potential to detect KNe independently of GW or GRB triggers, we expect high cadenced LSST observations to yield many serendipitous KN discoveries.

\software{
Astropy \citep{the_astropy_collaboration_2018_2556700},
Pandas \citep{reback2020pandas},
Matplotlib \citep{matplotlib},
Kowalski \citep{Duev2019},
lpipe \citep{Perley2019lpipe},
PostgreSQL\footnote{\url{https://www.postgresql.org/}},
PyRAF DBSP data reduction pipeline \citep{Bellm2016DBSP},
ztfquery \citep{mickael_rigault_2018_1345222}, SExtractor \citep{Bertin2010}, ForcePhotZTF \citep{Yao2019}.
}

\acknowledgements
\section*{Acknowledgments}

We thank P. Nugent for reviewing the manuscript and providing us with useful feedback.
We thank D. Perley for useful discussion.
Based on observations obtained with the Samuel Oschin 48-inch Telescope and the 60-inch Telescope at the Palomar Observatory as part of the Zwicky Transient Facility project, a  scientific  collaboration  among  the  California  Institute  of Technology,  the  Oskar  Klein  Centre,  the  Weizmann  Institute of Science, the University of Maryland, the University of Washington, Deutsches Elektronen-Synchrotron, the University of Wisconsin-Milwaukee, and the TANGO Program of the University System of Taiwan.  Further support is provided by the U.S. National Science Foundation under Grant No. AST-1440341. 

\noindent This work was supported by the GROWTH (Global Relay of Observatories Watching Transients Happen) project funded by the National Science Foundation under PIRE Grant No 1545949. GROWTH is a collaborative project among California Institute of Technology (USA), University of Maryland College Park (USA), University of Wisconsin Milwaukee (USA), Texas Tech University (USA), San Diego State University (USA), University of Washington (USA), Los Alamos National Laboratory (USA), Tokyo Institute of Technology (Japan), National Central University (Taiwan), Indian Institute of Astrophysics (India), Indian Institute of Technology Bombay (India), Weizmann Institute of Science (Israel), The Oskar Klein Centre at Stockholm University (Sweden), Humboldt University (Germany), Liverpool John Moores University (UK) and University of Sydney (Australia).

\noindent E.~C.~K., A.~S.~C., and A.~G.  acknowledge support from the G.R.E.A.T research environment funded by {\em Vetenskapsr\aa det},  the Swedish Research Council, project number 2016-06012.
E.~C.~K. also acknowledges support from The Wenner-Gren Foundations. 
M.~W.~C. acknowledges support from the National Science Foundation with grant number PHY-2010970.
D.~L.~K. is supported by NSF grant AST-1816492.

\noindent We thank A. Miller for the star/galaxy classification code development for ZTF.
This research has made use of the VizieR catalog access tool, CDS, Strasbourg, France (DOI: \texttt{10.26093/cds/vizier}). The original description of the VizieR service was published in A\&AS 143, 23.
This work has made use of data from the European Space Agency (ESA)
mission {\it Gaia} (\url{https://www.cosmos.esa.int/gaia}), processed by
the {\it Gaia} Data Processing and Analysis Consortium (DPAC,
\url{https://www.cosmos.esa.int/web/gaia/dpac/consortium}). Funding
for the DPAC has been provided by national institutions, in particular
the institutions participating in the {\it Gaia} Multilateral Agreement.

\bibliography{references, software}
\bibliographystyle{apj}

\begin{appendix}

\section{Kilonova candidates {\it post-mortem}}
\label{appendix: post-mortem}
We discuss here individual candidates. The main features that we consider are 1) variable catalogs cross-matches; 2) possible CLU galaxy cross-match, or presence of a bright nearby galaxy; 3) past variability in survey data other than ZTF; 4) color evolution.

\begin{itemize}

\item ZTF18aazjzed -- Reported to Transient Name Server (TNS) as AT2018ltl, it was catalogued as a cataclysmic variable. There are no PS1 pre-detections, but a faint source is visible in the ZTF reference image and there are recent ZTF detections (after the end date of the experiment). In addition, the color evolution is incompatible with KN models (Figure\,\ref{fig: gr_models_and_points}).

\item ZTF19abzwbxy -- The transient is located 86.5\,kpc from a CLU galaxy 67.2\,Mpc away. However, it is coincindent with a faint source present in the PS1 catalog, making an association with the CLU galaxy unlikely. In addition, the source is catalogued as variable in the AAVSO International Variable Star Index, possibly being a quasi-stellar object \citep[QSO;][]{Brescia2015}. 

\item ZTF18achqdzh -- Nuclear source located in a galaxy with spectroscopic redshift $z=0.0461$, so at peak the transient is at $M_r < -17$, probably too bright to be a KN \citep{Kasliwal2020arXiv}. The host was classified as a low-ionization nuclear emission-line region (LINER) and can harbor an AGN \citep{Cavuoti2014}. Sporadic faint detections in ZTF forced photometry suggest the emission to be due to nuclear variability.

\item ZTF20aaouvjn -- Reported to TNS as AT2020lru. Located 76.2\,kpc (388$''$) from a galaxy 40.5\,Mpc away. About 2.5$''$ from a star, most likely to be the source if it has large proper motion. No other sources are present nearby and there are no PS1 pre-detections. The color evolution is incompatible with KN models (Figure\,\ref{fig: gr_models_and_points}).

\item ZTF19abqtcob -- Reported to TNS as AT2019aadh, it is located 55\,kpc (1313$''$) from a CLU galaxy 8.6\,Mpc away. If the host is indeed the nearby CLU galaxy, this would imply an absolute magnitude of $M \sim -10$ at peak, consistent with a classical nova. The low Galactic latitude of $b_{\rm gal} = -10.0$, and the lack of any other possible host nearby make it likely for ZTF19abqtcob to be a foreground Galactic CV. 

\item ZTF18abyzkeq -- Hostless source with no PS1 pre-detections, ZTF18abyzkeq is classified as a variable star in the AAVSO International Variable Star Index. As expected, the color evolution is incompatible with KN models (Figure\,\ref{fig: gr_models_and_points}).

\item ZTF19aayhxlo -- Reported to TNS as AT2019aada, ZTF19aayhxlo appears to be a hostless source with no PS1 pre-detections.  The color evolution is incompatible with KN models (Figure\,\ref{fig: gr_models_and_points}). Recent ZTF detections (after the end date of the experiment) suggests the source to be a variable object.

\item ZTF19acecsfi -- Faint PS1 and SDSS pre-detection within $1''$. Deep Legacy Survey images suggest the presence of an underlying blue source deemed as stellar in Legacy Survey DR8 catalog \citep{Dey2019}. Despite the high Galactic latitude $b_{\rm{gal}} = 30.3$, we consider the source to be likely a CV.

\item ZTF19aabgebm -- Reported to TNS as AT2019aacx, ZTF19aabgebm is the (already known) optical afterglow of long GRB~190106A \citep{Sonbas2019GCN_GRB190106A, Yurkov2019GCN_GRB190106A} at redshift $z = 1.859$ \citep{Schady2019GCN_GRB190106A}. We note that its color and color evolution are barely compatible with KN models for NSBH mergers (Figure\,\ref{fig: gr_models_and_points})

\item ZTF20aahcrjn --  Reported to TNS as AT2020atp. A faint source is present at the candidate location in the ZTF reference image. There are no PS1 pre-detections, but forced photometry revealed another possible outburst in April 2018, suggesting this to be a Galactic variable.  Moreover, the color evolution is incompatible with KN models (Figure\,\ref{fig: gr_models_and_points}).

\item ZTF18abuzpri -- Reported to TNS as AT2018ltn. A prominent source is present in the ZTF images, likely consituted of 2 stars (Legacy Survey DR8). Although there is no activity recorded in the PS1 catalog, the likely stellarity and the significantly blue color $g-r = -0.73 \pm 0.18$ makes it unlikely for ZTF18abuzpri to be a KN. 

\item ZTF19abqiwjq -- Reported to TNS as AT2019aade, it is located in a crowded field at low Galactic latitude $b_{Gal} = -5.9$\,deg. A faint source is present in the reference image at the location of the candidate. PS1 detected this source multiple times in different filters and the flux in these detections has increased significantly (more than 8 times), indicating past activity. We therefore rule out ZTF19abqiwjq as a KN candidate. 

\item ZTF19abqneae -- Reported to TNS as AT2019aadd, it is located in a crowded stellar field at $b_{Gal} = -11.1 $\,deg. The candidate has multiple PS1 detections on different filters and the flux in these detections has doubled, showing activity in the past.

\item ZTF19acbtthv -- Reported to TNS as AT2019aacw. Hostless source, however PS1 pre-detections are present within 2$''$ and recent additional ZTF detections after the nominal end of the experiment, indicating repeated activity. Moreover, the color evolution is incompatible with KN models (Figure\,\ref{fig: gr_models_and_points}).

\item ZTF19acszwgx -- Previously reported to TNS by ATLAS team as AT2016hnn. An underlying faint source is visible in the ZTF reference image, with multiple PS1 detections within $1''$. The $g$-band PS1 flux varies ~30\% thus it was active in the past and a possible re-brightening could be detected in recent ZTF data.

\item ZTF19abcputm -- Reported to TNS as AT2019aacy. A faint source is present in the ZTF reference image, and it has multiple PS1 detection within $1''$. The r-band PS1 flux varies 300\% thus it was active in the past.

\item ZTF19abogdfr -- Reported to TNS as AT2019aacz, the source shows multiple PS1 detections within $1''$. The $g$-band PS1 flux varies by a factor of 80, thus it was active in the past. 

\item ZTF19aapbbde -- The source shows an underlying blue point-like source. ZTF19aapbbde has numerous PS1 detections within 1$''$, with flux measurements suggesting $\sim 50\%$ increase in its brightness. This is grater than what would be expected within the PSF flux 3-$\sigma$ error, suggesting previous activity.
The source shows dramatic reddening at a 3-4 days phase, which is incompatible with the expectations from KN models (Figure\,\ref{fig: gr_models_and_points}).

\item ZTF19abcpiag -- Previously reported by the Gaia team to TNS as AT2017dfl, this source is a catalogued variable star listed in the AAVSO International Variable Star Index. The transient presents an extremely rapid $r$-band light curve, but no color information is available.

\item ZTF18acfmhrt -- Reported to TNS as AT2018ltm, the source appears to be hostless with no pre-detections in PS1. An underlying faint blue source is visible in PS1 images. The color evolution is incompatible with KN models (Figure\,\ref{fig: gr_models_and_points}).

\item ZTF19abxwwmr -- Hostless and with no PS1 pre-detections. However, the color evolution is incompatible with KN models (Figure\,\ref{fig: gr_models_and_points}).

\item ZTF19aanhtzz -- Reported to TNS as AT2019aacu. The nearest source is $3.5''$ away and it is probably stellar (SDSS, Legacy Survey DR8). This source presents a number of interesting features, including red color, rapid evolution, and high Galactic latitude of $b_{\rm gal} = 59.2$\,deg. The lack of an apparent host galaxy suggests the source to be either a luminous cosmological transient (the afterglow of a GRB that went undetected by gamma-ray telescopes?) or some type of Galactic variable, rather than a kilonova. However, the red color disfavors the Galactic scenario and a conclusive answer regarding the nature of ZTF19aanhtzz/AT2019aacu is yet to be found. 

\item ZTF19abudvoz -- Previously reported to TNS by the ATLAS team as AT2016ayj in 2016. Apparently orphan, it has two PS1 pre-detections within $1''$. In addition, the light curve is also unlike what is expected for KNe, with no color evolution during the rise time and with a $g$-band decay of $\sim 1$\,mag\,day$^{-1}$.

\item ZTF18acsjqjd -- The source shows pre-activity in PS1 images and therefore we exclude that it can be a KN.

\end{itemize}

\section{Candidates fading rates}
\label{appendix: evolution rates}

Here we present the fading rates (mag per day) for the candidates that passed our selection criteria and quality checks. Specifically, we show for each band the index obtained by performing a linear fit from the brightest points light curves built with alerts, forced photometry, and nightly stacked forced photometry.

\begin{table*}
    \centering
    \begin{tabular}{lccccccccc}
    \hline \hline
Name & $\alpha_g$ & $\alpha_r$ & $\alpha_i$ & forced $\alpha_g$ & forced $\alpha_r$ & forced $\alpha_i$ & stack $\alpha_g$ & stack $\alpha_r$ & stack $\alpha_i$ \\
\hline
ZTF18aazjzed & 1.30 & 1.29 & nan & 1.29 & 1.27 & nan & nan & nan & nan \\
ZTF19abzwbxy & 0.82 & nan & nan & 0.82 & nan & nan & 0.82 & nan & nan \\
ZTF18achqdzh & nan & nan & nan & nan & 0.51 & nan & nan & nan & nan \\
ZTF20aaouvjn & nan & nan & nan & nan & nan & nan & 1.49 & 1.74 & nan \\
ZTF19abqtcob & nan & 0.39 & nan & nan & 0.43 & nan & nan & nan & nan \\
ZTF18abyzkeq & 0.85 & 0.45 & nan & 1.29 & 0.43 & nan & nan & 0.43 & nan \\
ZTF19aayhxlo & nan & nan & nan & 1.00 & 0.97 & nan & 1.00 & nan & nan \\
ZTF19acecsfi & 0.98 & nan & nan & 1.03 & nan & nan & 1.04 & nan & nan \\
ZTF19aabgebm & nan & nan & nan & 1.24 & 1.23 & nan & 1.24 & 1.24 & nan \\
ZTF20aahcrjn & 0.85 & 0.66 & nan & 0.89 & 0.63 & nan & 0.89 & 0.63 & nan \\
ZTF18abuzpri & nan & 8.89 & nan & nan & 9.09 & nan & nan & 3.50 & nan \\
ZTF19abqiwjq & nan & nan & nan & nan & 0.86 & nan & nan & 0.85 & nan \\
ZTF19abqneae & nan & nan & nan & nan & 1.10 & nan & nan & 1.05 & nan \\
ZTF19acbtthv & 1.44 & 1.52 & nan & 1.47 & 1.51 & nan & 1.48 & 1.51 & nan \\
ZTF19acszwgx & nan & 0.52 & nan & nan & 0.52 & nan & nan & 0.53 & nan \\
ZTF19abcputm & nan & 0.87 & nan & nan & 0.86 & nan & nan & 0.86 & nan \\
ZTF19abogdfr & nan & 0.63 & nan & nan & 0.62 & nan & nan & 0.64 & nan \\
ZTF19aapbbde & 1.11 & 0.50 & nan & 1.18 & 0.40 & nan & 1.18 & 0.41 & nan \\
ZTF19abcpiag & nan & nan & nan & nan & 1.14 & nan & nan & 1.23 & nan \\
ZTF18acfmhrt & 1.00 & nan & nan & 0.96 & 1.26 & nan & 0.96 & 1.26 & nan \\
ZTF19abxwwmr & nan & nan & nan & 0.89 & 1.05 & nan & 0.87 & 1.06 & nan \\
ZTF19aanhtzz & nan & nan & nan & nan & 1.25 & nan & nan & nan & nan \\
ZTF19abudvoz & nan & nan & nan & 0.91 & nan & nan & nan & nan & nan \\
ZTF18acsjqjd & nan & 0.42 & nan & nan & 0.42 & nan & nan & nan & nan \\
    \hline
    \end{tabular}
    \label{table: evolution rates}
\end{table*}

\clearpage

\section{kilonova rate grid for top-hat models}
\label{appendix: rates grid tophat}

\begin{longtable}{rrrr}
\toprule
  M &  $R$ (n=1) &  $R$ (n=2) &  $R$ (n=3) \\
     & [Gpc$^{-3}$\,yr$^{-1}$] &  [Gpc$^{-3}$\,yr$^{-1}$] &  [Gpc$^{-3}$\,yr$^{-1}$] \\
\midrule
-17.0 &           125.0 &           266.0 &           323.0 \\
-16.5 &           229.0 &           563.0 &           705.0 \\
-16.0 &           404.0 &           949.0 &          1162.0 \\
-15.5 &          1041.0 &          2238.0 &          2586.0 \\
-15.0 &          1595.0 &          3488.0 &          4285.0 \\
-14.5 &          2830.0 &          6521.0 &          7500.0 \\
-14.0 &          8571.0 &             NaN &             NaN \\
-13.5 &         10000.0 &             NaN &             NaN \\
-13.0 &         15000.0 &         15000.0 &         15000.0 \\
-12.5 &             NaN &             NaN &             NaN \\
-12.0 &             NaN &             NaN &             NaN \\
\bottomrule
\end{longtable}

\section{kilonova rate grid for linear decay models}
\label{appendix: rates grid linear}

The following table presents results for a grid of linear decay models used to inject synthetic light curves in \texttt{simsurvey}. The columns correspond to the starting absolute magnitude, the decay rate, and the rates corresponding to 1, 2, 3 detections ($5\sigma$) in ZTF at 95\% confidence.

\begin{longtable}{rrrrr}
\toprule
  M &  $\alpha$ &  $R$ (n=1) &  $R$ (n=2) &  $R$ (n=3) \\
     &  [mag\,day$^{-1}$] & [Gpc$^{-3}$\,yr$^{-1}$] &  [Gpc$^{-3}$\,yr$^{-1}$] &  [Gpc$^{-3}$\,yr$^{-1}$] \\
\midrule
-17.5 &   0.3 &             82 &          141 &          184 \\
-17.5 &   0.4 &             85 &          173 &          226 \\
-17.5 &   0.5 &             99 &          200 &          251 \\
-17.5 &   0.6 &            112 &          242 &          335 \\
-17.5 &   0.7 &            140 &          306 &          379 \\
-17.5 &   0.8 &            138 &          314 &          413 \\
-17.5 &   0.9 &            135 &          340 &          447 \\
-17.5 &   1.0 &            163 &          500 &          674 \\
-17.5 &   1.1 &            181 &          545 &          714 \\
-17.5 &   1.2 &            168 &          540 &          689 \\
-17.5 &   1.3 &            187 &          594 &          789 \\
-17.5 &   1.4 &            200 &          674 &          983 \\
-17.5 &   1.5 &            205 &          750 &          967 \\
-17.0 &   0.3 &            154 &          288 &          377 \\
-17.0 &   0.4 &            179 &          335 &          444 \\
-17.0 &   0.5 &            173 &          363 &          465 \\
-17.0 &   0.6 &            232 &          535 &          697 \\
-17.0 &   0.7 &            250 &          600 &          895 \\
-17.0 &   0.8 &            272 &          722 &          967 \\
-17.0 &   0.9 &            287 &          740 &         1016 \\
-17.0 &   1.0 &            317 &          895 &         1395 \\
-17.0 &   1.1 &            315 &         1016 &         1276 \\
-17.0 &   1.2 &            344 &         1250 &         1874 \\
-17.0 &   1.3 &            375 &         1333 &         1666 \\
-17.0 &   1.4 &            382 &         1250 &         1666 \\
-17.0 &   1.5 &            441 &         1764 &         2500 \\
-16.5 &   0.3 &            304 &          508 &          645 \\
-16.5 &   0.4 &            408 &          800 &         1090 \\
-16.5 &   0.5 &            422 &          740 &         1052 \\
-16.5 &   0.6 &            397 &          952 &         1132 \\
-16.5 &   0.7 &            447 &          937 &         1363 \\
-16.5 &   0.8 &            545 &         1463 &         1874 \\
-16.5 &   0.9 &            576 &         1428 &         1874 \\
-16.5 &   1.0 &            550 &         1764 &         2142 \\
-16.5 &   1.1 &            530 &         1621 &         2000 \\
-16.5 &   1.2 &            666 &         2142 &         2500 \\
-16.5 &   1.3 &            594 &         2000 &         2857 \\
-16.5 &   1.4 &            759 &         2727 &         4000 \\
-16.5 &   1.5 &            789 &         2500 &         3333 \\
-16.0 &   0.3 &            504 &          857 &         1071 \\
-16.0 &   0.4 &            779 &         1304 &         1874 \\
-16.0 &   0.5 &            645 &         1395 &         2000 \\
-16.0 &   0.6 &            833 &         1500 &         2068 \\
-16.0 &   0.7 &            789 &         1935 &         3333 \\
-16.0 &   0.8 &           1395 &         3000 &         4285 \\
-16.0 &   0.9 &            983 &         2400 &         3157 \\
-16.0 &   1.0 &           1276 &         5000 &         6666 \\
-16.0 &   1.1 &           1428 &         4285 &         5000 \\
-16.0 &   1.2 &           1395 &         4615 &         6000 \\
-16.0 &   1.3 &           1304 &         5000 &         6000 \\
-16.0 &   1.4 &           1304 &         3333 &         4000 \\
-16.0 &   1.5 &           1304 &         5000 &         6000 \\
-15.5 &   0.3 &           1276 &         2222 &         3000 \\
-15.5 &   0.4 &           1200 &         2500 &         3333 \\
-15.5 &   0.5 &           1935 &         5000 &         5454 \\
-15.5 &   0.6 &           1935 &         5454 &         6666 \\
-15.5 &   0.7 &           2000 &         5000 &         6000 \\
-15.5 &   0.8 &           2222 &         5000 &         5454 \\
-15.5 &   0.9 &           2400 &         6666 &        10000 \\
-15.5 &   1.0 &           1764 &         8571 &        12000 \\
-15.5 &   1.1 &           2142 &         5000 &         7500 \\
-15.5 &   1.2 &           2727 &        12000 &        15000 \\
-15.5 &   1.3 &           3750 &            nan &            nan \\
-15.5 &   1.4 &           4615 &        20000 &            nan \\
-15.5 &   1.5 &           6000 &            nan &            nan \\
-15.0 &   0.3 &           3333 &         7500 &        12000 \\
-15.0 &   0.4 &           2222 &         4285 &         4615 \\
-15.0 &   0.5 &           3000 &         5000 &         5454 \\
-15.0 &   0.6 &           2222 &         5000 &         6666 \\
-15.0 &   0.7 &           2142 &         4615 &         5000 \\
-15.0 &   0.8 &           4285 &        12000 &        12000 \\
-15.0 &   0.9 &           5454 &        20000 &        20000 \\
-15.0 &   1.0 &           3000 &        10000 &            nan \\
-15.0 &   1.1 &           4615 &        10000 &        12000 \\
-15.0 &   1.2 &           3750 &        12000 &        15000 \\
-15.0 &   1.3 &           3750 &        15000 &        20000 \\
-15.0 &   1.4 &           7500 &            nan &            nan \\
-15.0 &   1.5 &          10000 &        10000 &            nan \\
-14.5 &   0.3 &           5000 &         6000 &         6000 \\
-14.5 &   0.4 &           5454 &        10000 &        10000 \\
-14.5 &   0.5 &           4285 &        10000 &        12000 \\
-14.5 &   0.6 &           7500 &        15000 &        20000 \\
-14.5 &   0.7 &           8571 &        15000 &        15000 \\
-14.5 &   0.8 &           7500 &        15000 &        15000 \\
-14.5 &   0.9 &          15000 &        20000 &            nan \\
-14.5 &   1.0 &           8571 &            nan &            nan \\
-14.5 &   1.1 &          10000 &            nan &            nan \\
-14.5 &   1.2 &           8571 &        15000 &        15000 \\
-14.5 &   1.3 &           7500 &            nan &            nan \\
-14.5 &   1.4 &          15000 &            nan &            nan \\
-14.5 &   1.5 &          10000 &        10000 &        10000 \\
\bottomrule
\end{longtable}

\section{kilonova rate for best-fit model AT2017gfo}
\label{appendix: rates AT2017gfo}

\begin{longtable}{rrrr}
\toprule
  viewing angle &  $R$ (n=1) &  $R$ (n=2) &  $R$ (n=3) \\
    $[$deg$]$ & [Gpc$^{-3}$\,yr$^{-1}$] &  [Gpc$^{-3}$\,yr$^{-1}$] &  [Gpc$^{-3}$\,yr$^{-1}$] \\
\midrule
     fixed 20$^{\circ}$ &   613 &  1775 &  2400 \\
     uniform distribution in cosine &  1273 &  4029 &  5510 \\
\bottomrule
\end{longtable}

\section{kilonova rate ejecta mass BNS grid}
\label{appendix: rates mej grid}

\begin{longtable}{rrrrr}
\toprule
  $m_{dyn}$     & $m_{win}$     &  $R$ (n=1) &  $R$ (n=2) &  $R$ (n=3) \\
   $[$M$_{\odot}]$ &  $[$M$_{\odot}]$ & [Gpc$^{-3}$\,yr$^{-1}$] &  [Gpc$^{-3}$\,yr$^{-1}$] &  [Gpc$^{-3}$\,yr$^{-1}$] \\
\midrule
 0.001 &  0.01 &            5000 &           12000 &           14117 \\
 0.001 &  0.03 &            2876 &            6562 &            9545 \\
 0.001 &  0.05 &            2142 &            5384 &            8076 \\
 0.001 &  0.07 &            1500 &            2957 &            4117 \\
 0.001 &  0.09 &            1478 &            3230 &            4200 \\
 0.001 &  0.11 &            1354 &            2727 &            3620 \\
 0.001 &  0.13 &            1272 &            3230 &            4772 \\
 0.005 &  0.01 &            2000 &            6000 &           11250 \\
 0.005 &  0.03 &            1153 &            3396 &            4390 \\
 0.005 &  0.05 &             845 &            2278 &            3050 \\
 0.005 &  0.07 &             711 &            1800 &            2368 \\
 0.005 &  0.09 &             666 &            1636 &            2222 \\
 0.005 &  0.11 &             538 &            1304 &            1682 \\
 0.005 &  0.13 &             517 &            1200 &            1636 \\
 0.010 &  0.01 &            1440 &            4090 &            6428 \\
 0.010 &  0.03 &             694 &            2000 &            2535 \\
 0.010 &  0.05 &             520 &            1538 &            2117 \\
 0.010 &  0.07 &             491 &            1525 &            1978 \\
 0.010 &  0.09 &             377 &             913 &            1276 \\
 0.010 &  0.11 &             378 &             957 &            1267 \\
 0.010 &  0.13 &             349 &             829 &            1132 \\
 0.020 &  0.01 &            1052 &            3103 &            3750 \\
 0.020 &  0.03 &             590 &            2168 &            3333 \\
 0.020 &  0.05 &             396 &            1129 &            1500 \\
 0.020 &  0.07 &             331 &             871 &            1250 \\
 0.020 &  0.09 &             269 &             711 &            1004 \\
 0.020 &  0.11 &             253 &             666 &             878 \\
 0.020 &  0.13 &             228 &             623 &             850 \\
\bottomrule
\end{longtable}

\section{kilonova rate ejecta mass NSBH grid}
\label{appendix: rates mej grid}

\begin{longtable}{rrrrr}
\toprule
  $m_{dyn}$     & $m_{win}$     &  $R$ (n=1) &  $R$ (n=2) &  $R$ (n=3) \\
   $[$M$_{\odot}]$ &  $[$M$_{\odot}]$ & [Gpc$^{-3}$\,yr$^{-1}$] &  [Gpc$^{-3}$\,yr$^{-1}$] &  [Gpc$^{-3}$\,yr$^{-1}$] \\
\midrule
  0.01 &  0.01 &            4384 &           11632 &           16285 \\
 0.01 &  0.02 &            3677 &           10555 &           14250 \\
 0.01 &  0.03 &            2226 &            5229 &            6867 \\
 0.01 &  0.04 &            2244 &            5876 &            8028 \\
 0.01 &  0.05 &            1821 &            4318 &            6195 \\
 0.01 &  0.06 &            1701 &            4318 &            5700 \\
 0.01 &  0.07 &            1503 &            3986 &            5000 \\
 0.01 &  0.08 &            1393 &            3220 &            4710 \\
 0.01 &  0.09 &            1212 &            2968 &            4285 \\
 0.02 &  0.01 &            3617 &            7846 &           10625 \\
 0.02 &  0.02 &            3128 &            8225 &           12439 \\
 0.02 &  0.03 &            2628 &            5862 &            7968 \\
 0.02 &  0.04 &            2056 &            4358 &            5730 \\
 0.02 &  0.05 &            1984 &            4473 &            6071 \\
 0.02 &  0.06 &            1634 &            3445 &            5257 \\
 0.02 &  0.07 &            1522 &            3669 &            5257 \\
 0.02 &  0.08 &            1370 &            2982 &            4080 \\
 0.02 &  0.09 &            1338 &            2931 &            4047 \\
 0.03 &  0.01 &            3355 &            9107 &           10625 \\
 0.03 &  0.02 &            2524 &            6219 &            8360 \\
 0.03 &  0.03 &            2383 &            5543 &            7727 \\
 0.03 &  0.04 &            2286 &            4766 &            6710 \\
 0.03 &  0.05 &            2023 &            4766 &            6144 \\
 0.03 &  0.06 &            1634 &            3333 &            4678 \\
 0.03 &  0.07 &            1522 &            3493 &            4636 \\
 0.03 &  0.08 &            1370 &            3072 &            4080 \\
 0.03 &  0.09 &            1256 &            2698 &            3617 \\
 0.04 &  0.01 &            2756 &            6000 &            7083 \\
 0.04 &  0.02 &            2428 &            5730 &            8095 \\
 0.04 &  0.03 &            2116 &            4146 &            5604 \\
 0.04 &  0.04 &            2207 &            4473 &            6800 \\
 0.04 &  0.05 &            2073 &            4553 &            6986 \\
 0.04 &  0.06 &            1694 &            3445 &            4811 \\
 0.04 &  0.07 &            1432 &            2897 &            3750 \\
 0.04 &  0.08 &            1495 &            3128 &            4722 \\
 0.04 &  0.09 &            1404 &            2771 &            4080 \\
 0.05 &  0.01 &            2451 &            4594 &            5862 \\
 0.05 &  0.02 &            2512 &            4766 &            6623 \\
 0.05 &  0.03 &            2081 &            4473 &            6538 \\
 0.05 &  0.04 &            2073 &            4146 &            5730 \\
 0.05 &  0.05 &            1969 &            3566 &            5151 \\
 0.05 &  0.06 &            1578 &            3333 &            4636 \\
 0.05 &  0.07 &            1495 &            2897 &            3805 \\
 0.05 &  0.08 &            1378 &            2475 &            3566 \\
 0.05 &  0.09 &            1297 &            2500 &            3227 \\
 0.06 &  0.01 &            2550 &            5151 &            7083 \\
 0.06 &  0.02 &            2256 &            4636 &            5862 \\
 0.06 &  0.03 &            2417 &            5368 &            7727 \\
 0.06 &  0.04 &            2000 &            3692 &            5052 \\
 0.06 &  0.05 &            1951 &            3749 &            5393 \\
 0.06 &  0.06 &            1454 &            2637 &            3692 \\
 0.06 &  0.07 &            1618 &            3191 &            4166 \\
 0.06 &  0.08 &            1496 &            2984 &            4042 \\
 0.06 &  0.09 &            1289 &            2467 &            3114 \\
 0.07 &  0.01 &            2368 &            5113 &            6428 \\
 0.07 &  0.02 &            2132 &            4054 &            5696 \\
 0.07 &  0.03 &            2261 &            5294 &            7031 \\
 0.07 &  0.04 &            1867 &            3629 &            5113 \\
 0.07 &  0.05 &            1679 &            3284 &            4368 \\
 0.07 &  0.06 &            1704 &            3600 &            5232 \\
 0.07 &  0.07 &            1685 &            3191 &            4205 \\
 0.07 &  0.08 &            1535 &            3169 &            4245 \\
 0.07 &  0.09 &            1319 &            2777 &            3719 \\
 0.08 &  0.01 &            1867 &            3879 &            5921 \\
 0.08 &  0.02 &            2083 &            4205 &            5921 \\
 0.08 &  0.03 &            2031 &            3939 &            6724 \\
 0.08 &  0.04 &            1788 &            4239 &            6190 \\
 0.08 &  0.05 &            1751 &            3453 &            4615 \\
 0.08 &  0.06 &            1777 &            3404 &            4485 \\
 0.08 &  0.07 &            1384 &            2472 &            3169 \\
 0.08 &  0.08 &            1465 &            2760 &            3750 \\
 0.08 &  0.09 &            1235 &            2346 &            3360 \\
 0.09 &  0.01 &            1925 &            3564 &            4615 \\
 0.09 &  0.02 &            1925 &            4235 &            6000 \\
 0.09 &  0.03 &            1836 &            3600 &            4337 \\
 0.09 &  0.04 &            2130 &            4615 &            6206 \\
 0.09 &  0.05 &            1825 &            3692 &            5217 \\
 0.09 &  0.06 &            1685 &            3082 &            3982 \\
 0.09 &  0.07 &            1555 &            3134 &            4285 \\
 0.09 &  0.08 &            1390 &            2709 &            3925 \\
 0.09 &  0.09 &            1220 &            2320 &            3157 \\
\bottomrule
\end{longtable}

\end{appendix}

\end{document}